\documentclass[trackchanges, preprint]{aastex631}
\PassOptionsToPackage{dvipsnames}{xcolor}
\pdfoutput=1
\usepackage{amsmath}
\usepackage{capt-of}   
\usepackage{empheq}
\usepackage{enumitem}
\usepackage{hyperref}
\usepackage{hhline}
\usepackage{numprint}
\usepackage{pifont}
\usepackage{xspace}
\usepackage{placeins}
\usepackage{graphicx}
\usepackage{mwe}
\usepackage{multirow}
\usepackage{tabularx,booktabs}
\usepackage{hhline}
\usepackage{booktabs}
\usepackage{changepage}
\usepackage{float}
\usepackage{booktabs} 
\usepackage{array} 

\newcommand{\totalprelim}{1346\xspace}
\newcommand{\gradeAprelim}{96\xspace}
\newcommand{\gradeBprelim}{326\xspace}
\newcommand{\gradeCprelim}{924\xspace}

\newcommand{\totalfinal}{1346\xspace}
\newcommand{\gradeAfinal}{146\xspace}
\newcommand{\gradeBfinal}{199\xspace}
\newcommand{\gradeCfinal}{1001\xspace}

\newcolumntype{Y}{>{\centering\arraybackslash}X}

\shorttitle{UNIONS Strong Lens Search}

\begin{document}

\title{Gravitational Lenses in UNIONS and Euclid (GLUE) I:\\ A Search for Strong Gravitational Lenses in UNIONS with Subaru, CFHT, and Pan-STARRS Data}

\submitjournal{ApJ}

\shorttitle{Gravitational Lenses in UNIONS \& Euclid}
\shortauthors{Storfer, Magnier et al.}

\correspondingauthor{Christopher J. Storfer}
\email{cstorfer@hawaii.edu}

\author[0000-0002-0385-0014]{Christopher J. Storfer}
\affiliation{Institute for Astronomy, University of Hawai`i, 2680 Woodlawn Drive, Honolulu, HI 96822, USA}
\affiliation{Physics Division, Lawrence Berkeley National Laboratory, 1 Cyclotron Road, Berkeley, CA, 94720}

\author[0000-0002-7965-2815]{Eugene A. Magnier}
\affiliation{Institute for Astronomy, University of Hawai`i, 2680 Woodlawn Drive, Honolulu, HI 96822, USA}

\author[0000-0001-8156-0330]{Xiaosheng Huang}
\affiliation{Department of Physics \& Astronomy, University of San Francisco, San Francisco, CA 94117-1080}

\author[0000-0001-5402-4647]{David Rubin}
\affiliation{Department of Physics and Astronomy, University of Hawai`i at Mānoa, Honolulu, Hawai`i 96822}
\affiliation{Physics Division, Lawrence Berkeley National Laboratory, 1 Cyclotron Road, Berkeley, CA, 94720}

\author[0000-0002-5042-5088]{David J. Schlegel}
\affiliation{Physics Division, Lawrence Berkeley National Laboratory, 1 Cyclotron Road, Berkeley, CA, 94720}

\author[0009-0005-0646-0395]{Saurav Banka}
\affiliation{Department of Electrical Engineering \& Computer Sciences, University of California, Berkeley, Berkeley, CA 94720}

\author[0000-0001-6965-7789]{Kenneth C. Chambers}
\affiliation{Institute for Astronomy, University of Hawai`i, 2680 Woodlawn Drive, Honolulu, HI 96822, USA}

\author[0000-0002-3263-8645 ]{Jean-Charles Cuillandre}
\affiliation{Universit\'e Paris-Saclay, Universit\'e Paris Cit\'e, CEA, CNRS, AIM, 91191, Gif-sur-Yvette, France }

\author[0000-0001-5486-2747]{Thomas de Boer}
\affiliation{Institute for Astronomy, University of Hawai`i, 2680 Woodlawn Drive, Honolulu, HI 96822, USA}

\author[0000-0002-5540-6935]{Rapha\"el  Gavazzi}
\affiliation{Laboratoire d'Astrophysique de Marseille, CNRS, Aix-Marseille Universit\'e, CNES, Marseille, France} \affiliation{Institut d'Astrophysique de Paris, UMR 7095, CNRS, and Sorbonne Universit\'e, 98 bis boulevard Arago, 75014 Paris}

\author[0000-0001-8221-8406]{Stephen Gwyn}
\affiliation{National Research Council Herzberg Astronomy and Astrophysics, 5071 West Saanich Road, Victoria, B.C., V8Z6M7, Canada}

\author[0000-0002-1437-3786]{Michael J. Hudson}  
\affiliation{Department of Physics and Astronomy, University of Waterloo, 200 University Avenue West, Waterloo, Ontario N2L 3G1, Canada}
\affiliation{Waterloo Centre for Astrophysics, University of Waterloo, Waterloo, Ontario N2L 3G1, Canada}
\affiliation{Perimeter Institute for Theoretical Physics, 31 Caroline St. North, Waterloo, ON N2L 2Y5, Canada}

\author[0000-0002-6639-6533]{Gregory S. H. Paek}  
\affiliation{Institute for Astronomy, University of Hawaii, 2680 Woodlawn Drive, Honolulu HI 96822}

\author[0009-0008-8718-0644]{Douglas Scott}  
\affiliation{Department of Physics and Astronomy, University of British Columbia, 6225 Agricultural Road, Vancouver, V6T 1Z1, Canada}


\begin{abstract}
We present the results of our pipeline for discovering strong gravitational lenses in the ongoing Ultraviolet Near-Infrared Optical Northern Survey (UNIONS). We successfully train the deep residual neural network (ResNet) based on  \texttt{CMU-Deeplens} architecture, which is designed to detect strong lenses in ground-based imaging surveys. We train on images of real strong lenses and deploy on a sample of 8 million galaxies in areas with full coverage in the \textit{g}, \textit{r}, and \textit{i} filters, the first multi-band search for strong gravitational lenses in UNIONS. Following human inspection and grading, we report the discovery of a total of \totalprelim new strong lens candidates of which \gradeAfinal are grade A, \gradeBfinal grade B, and \gradeCfinal grade C. Of these candidates, 283 have lens-galaxy spectroscopic redshifts from the Sloan Digital Sky Survey (SDSS) and an additional 297 from the Dark Energy Spectroscopic Instrument (DESI) Data Release 1 (DR1). We find 15 of these systems display evidence of both lens and source galaxy redshifts in spectral superposition.
We additionally report the spectroscopic confirmation of seven lensed sources in high-quality systems, all with $z>2.1$, using the Keck Near-Infrared Echelle Spectrograph (NIRES) and Gemini Near-Infrared Spectrograph (GNIRS).
\end{abstract}
\section{Introduction} \label{sec:intro}

The nature of strong lensing enables it to act as a powerful probe for a range of astrophysical and cosmological studies. Strong-lensing observables are directly determined by the total mass and mass profile of the lens, thus they can be used as a tool to study massive galaxies and clusters, as well as granting insight to the distribution of dark matter within them \citep[e.g.,][]{kochanek1991a,bolton2006a,huang2009a}. In the same way, the multiple images and arcs produced by strong lensing are highly sensitive to perturbations by low-mass dark matter halos ($<10^9\,M\odot$), either as subhalos within the lens \citep[e.g.,][]{vegetti2009a} or along the line of sight \citep[e.g.,][]{cagansengul2021a}. Such studies can provide valuable tests of the cold dark matter (CDM) model beyond these coming from the local Universe.

As a cosmological probe, strong lensing has been shown to produce competitive measurements of the Hubble constant \citep[$H_0$; e.g.,][]{wong2019a} by utilizing lensed images with measurable time delays and variable sources such as quasars (QSOs) or supernovae (SNe). This method of measurement is complementary to early Universe probes such as the cosmic microwave background \citep{planck2020a} and is independent of local measurements, such as the cosmic distance ladder that uses Cepheids and type Ia SNe \citep[][]{riess2021a}. Strong lensing may also serve as a  geometric probe of cosmology through the use of lensing systems containing several multiply-imaged sources at different redshifts \citep[e.g.,][]{collett2014,caminha2022,sheu2024a}. Ratios of distances between observer, lens, and each of the multiple sources are highly sensitive to cosmology, especially the dark energy equation of state \citep[][]{linder2016,sharma2022a}. 

The rapid discovery of numerous new strong lenses in recent years has allowed for initial explorations of the full range of strong-lensing science applications. Early searches for strong lenses were conducted by manually inspecting images \citep[e.g.,][]{faure2008a} or through the use of analytical algorithms \citep[][]{more2012a,gavazzi2014a}. However, with the growing prevalence of deep, wide-field imaging surveys, more efficient search methods have become increasingly important. Within the last decade, the introduction of machine learning, and in particular convolutional neural networks (CNNs), to the search for strong lenses has been transformative \citep[e.g.,][]{metcalf2018a,lanusse2018a}, growing the number of strong lens candidates from $\mathcal{O}(10^2)$ to $\mathcal{O}(10^4)$, despite their intrinsic rarity. 

A large majority of strong gravitational lens candidates have been discovered at declinations below 32$^{\circ}$, for example in the Dark Energy Survey \citep[DES; ][]{jacobs2019b,gonzalez2025a}, the Dark Energy Spectroscopic Instrument Legacy Surveys \citep[DESI LS;][]{huang2020a, huang2021a, stein2021a,storfer2024a}, other Dark Energy Camera (DECam) surveys \citep[][]{holwerda2022a,zaborowski2023a}, and the Kilo Degree Survey \citep[KiDS;][]{petrillo2019a,nagam2024a}. To date, the northern sky has only sparsely been searched due to the lack of surveys with the necessary combination of depth, area, and image quality, all of which are critical in the search for strong lenses \citep[][]{collett2015a}. Searches in the northern portion 
($\delta\gtrsim32^{\circ}$) 
of DESI LS have proven to be successful yet candidates are reported at a lower density per deg$^2$ when compared with the Hyper Suprime-Cam Strategic Survey Program \citep[HSC-SSP; e.g.,][]{sonnenfeld2018a,shu2022a}. 
Additionally, \citet{savary2021a} reports 104 new candidates in $2500$~deg.$^2$ of the Canada-France Imaging Survey (CFIS) $r$-band data. CFIS is a component of the relatively new northern sky survey, the Ultraviolet Near Infrared Optical Northern Survey \citep[UNIONS;][]{gwyn2025a}. 

UNIONS shows great promise for new strong lens discovery and thus in this publication we present the results of our search for strong gravitational lens candidates in UNIONS. We provide an overview of the observations in \S\ref{sec:unions}. The construction of our training sample and the training of the neural network model is described in \S\ref{sec:search}. In \S\ref{sec:results}
 we report our \totalprelim new strong lens candidates. We discuss and contextualize our discoveries and show a number of fully confirmed strong lenses in \S\ref{sec:discussion} and conclude in \S\ref{sec:conc}.
\section{Observations and Data} \label{sec:unions}
\subsection{UNIONS}
The Ultraviolet Near Infrared Optical Northern Survey \citep[UNIONS;][]{gwyn2025a} is an ongoing wide-field sky survey. 
UNIONS is a compilation of surveys from four telescopes (three observatories) in five filters, $ugriz$. The Canada-France-Hawai`i Telescope (CFHT; 3.6\,m) provides the \textit{u-} and \textit{r-} band data for the Canada France Imaging Survey (CFIS). The Subaru Telescope (8.2\,m) provides the \textit{g}-band and portions of the \textit{z}-band data for the Waterloo-Hawai`i-IfA G-band Survey (WHIGS) and Wide Imaging with Subaru HSC of the Euclid Sky (WISHES), respectively. The Panoramic Survey Telescope And Rapid Response System (Pan-STARRS; 2$\times$1.8\,m) provides data for the \textit{i}-band, as well as additional \textit{z}-band coverage to complement the WISHES area. Observations by both CFHT and Subaru provide exquisite image quality, with reported median seeing of 0.71$^{\prime\prime}$ and 0.90$^{\prime\prime}$, respectively. As reported in \citet{gwyn2025a}, the 10$\sigma$ point-source depth (with a 2$^{\prime\prime}$ diameter) in the $ugriz$ filters are 23.7, 24.5, 24.2, 23.8, and 23.3 mags, respectively. We convert this to 5$\sigma$ limiting magnitudes of 22.95, 23.75, 23.45, 23.05, and 22.55 for the $ugriz$ filters. Throughout the remainder of this work we will reference the 5$\sigma$ limiting magnitude.
The core survey region covers 6250 deg$^2$ of the northern extragalactic sky, with a majority of the area in the north Galactic cap (NGC) at $\delta>30^{\circ}$, along with additional components in the NGC at 
 $\delta<30^{\circ}$ and the south Galactic cap (SGC). The survey is segmented into ``skycells'' which are 0.5$^{\circ}$$\times$0.5$^{\circ}$ images. Each sub-survey is resampled such that their pixel scales match the highest resolution images (HSC; $0.17^{\prime\prime}$ per pixel) and put on the same tesselation (CFIS.V0).
For additional details on individual survey coverage, depth, and image quality see tables 1, 3, and 4 from \citet{gwyn2025a}. 

In this work we make use of the $gri$ data available as of early 2024 in the NGC at $\delta>30^{\circ}$. Unless otherwise specified, all descriptions of the survey will reference this dataset. Table~\ref{tab:UNIONS-coverage} shows the approximate survey area and Figure~\ref{fig:map} shows the survey footprint for the individual filters and their combined regions. Figure~\ref{fig:dmap} shows the 5$\sigma$ point-spread function (PSF) depth for each of the $g$ $r$, and $i$ filters.

UNIONS serves as the ground-based counterpart to the northern portion of the \textit{Euclid} space telescope survey \citep[][]{mellier2024a}. \textit{Euclid} will observe 14,000 deg$^2$, delivering observations in the visible (VIS; approximately SDSS $riz$ filters) with point-spread function of 0.16 arcsec and a pixel scale of 0.1 arcsec per pixel. Additionally, \textit{Euclid} will provide lower-resolution infrared (IR) imaging at 0.3 arcsec per pixel in each of the $Y$, $J$, and $H$ bands via the Near-infrared spectrometer and photometer (NISP).  
UNIONS will provide multi-wavelength ground-based photometry and photometric redshifts for \textit{Euclid} while also delivering on independent science goals such as probing the properties of dark matter, studying large-scale structure, and exploring Milky Way science. Due to the survey area, depth, and image quality, UNIONS provides unparalleled coverage of the northern hemisphere from the ground. 

\begin{table}[h]
    \centering
    \caption{UNIONS Coverage Area}\label{tab:UNIONS-coverage}
    \begin{minipage}{\textwidth}
        \hspace{-1.6cm}
        \centering
        \begin{tabular}{lcccc}
            \hline
            Filter & \textit{g} & \textit{r} & \textit{i} & \textit{gri} \\
            \hline
            Area [deg$^2$] & 4700 & 4600 & 5500 & 4100 \\
            \hline
        \end{tabular}
        
        \vspace{0.2cm}
        
        \parbox{0.6\textwidth}{\footnotesize NOTE--There are available \textit{i}-band data in the south Galactic cap (SGC) from an earlier processing. We opt to exclude these, as well as the \textit{g-} and \textit{r-}band data in that region for this work. All values are approximations of total survey area.}
    \end{minipage}
\end{table}
\vspace{-0.4cm}
\begin{minipage}{0.9\textwidth}
\makebox[\textwidth][c]{
  \includegraphics[keepaspectratio=true,scale=0.7]{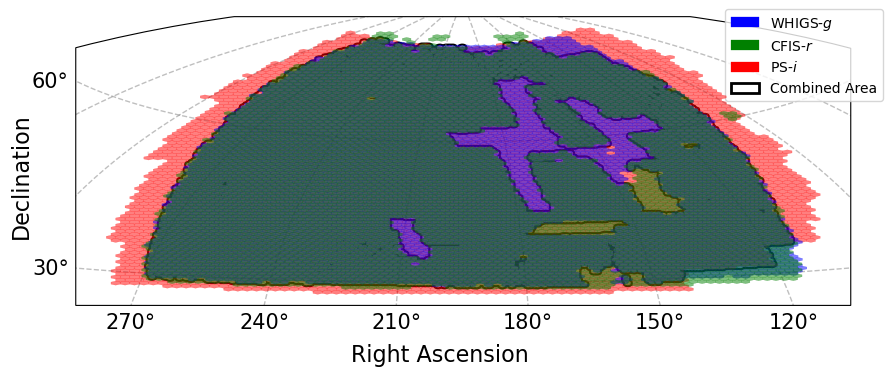}}
    \captionof{figure}{\footnotesize Footprint of the UNIONS \textit{gri} data used in this work in an equal-area Aitoff projection in equatorial coordinates with hexagonal binning. The WHIGS \textit{g} band is shown in blue, CFIS \textit{r} band in green, and PS \textit{i} band in red. There is  4100 deg$^2$ of \textit{gri} coverage the north Galactic cap (NGC).}
\label{fig:map}
\end{minipage}

\vspace{0.5cm}
\begin{minipage}{0.95\textwidth}
\makebox[\textwidth][c]{
  \includegraphics[keepaspectratio=true,scale=0.25]{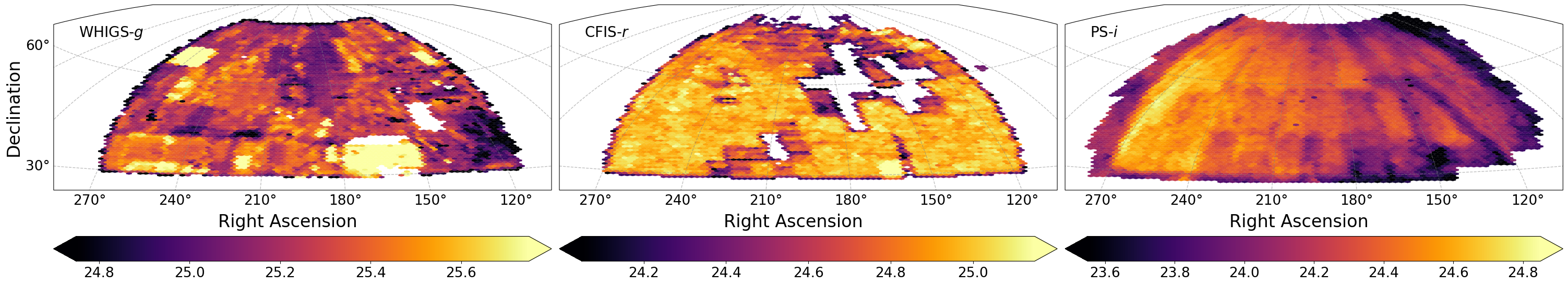}}
    \captionof{figure}{\footnotesize Depth maps for the individual surveys that make up the $gri$ component of UNIONS on an equal-area Aitoff projection in equatorial coordinates with hexagonal binning. In each panel we show the average 5$\sigma$ PSF depth for the survey with the corresponding colorbars shown below. From left to right; WHIGS $g$ band; CFIS $r$ band; and PS $i$ band. These have a median depths of 25.29, 24.92, and 24.23, respectively.}
\label{fig:dmap}
\end{minipage}



\subsection{\textit{The Tractor} Catalog}
\label{sec:tractor}
At the time of this work, UNIONS does not have a comprehensive object catalog which models objects in all filters simultaneously. For this reason we opt to use \textit{the Tractor} \citep[][]{lang2016a} object catalogs from the DESI LS Data Release 9 which covers the entirety of the UNIONS footprint.
\textit{The Tractor} is a forward modeling algorithm that performs probabilistic astronomical source detection and classification to construct source catalogs. 
Source extraction is done on pixel-level data, taking as input the individual images from multiple exposures in multiple bands, with different seeing in each.\footnote{For more details, see \href{https://www.legacysurvey.org/dr9/description/}{https://www.legacysurvey.org/dr9/description/}} 
\textit{The Tractor} classifies sources as the best of five light profile types: point source (``{\tt PSF}''); round exponential (``{\tt REX}''); de~Vaucouleurs ({\tt DEV}, $n_\text{s}=4$); exponential ({\tt EXP}; $n_\text{s}=1$); or S\'ersic ({\tt SER}). The latter four are all galaxy types, with {\tt REX}, {\tt DEV}, and {\tt SER} often being elliptical galaxies. \citet{huang2020a}, \citet{huang2021a}, and \citet{storfer2024a} all make use of \textit{Tractor} catalogs from their respective data releases for their strong lens searches in the DESI LS. In this work we make use of the exact same catalog as \citet{storfer2024a}. They apply an additional cut on the $z$-band magnitudes ($<$20), citing diminishing returns in visual inspection beyond this cut. We do not make use of the UNIONS $z$-band data from WISHES or PS in this work due to a combination of limited coverage and incomplete availability of fully-processed images at the start of our investigation (early 2024).  Nonetheless, we maintain this magnitude cut for the DESI LS catalog used in this work. We also apply a cut to the object types, limiting the catalog to primarily elliptical galaxies (exclusion of {\tt PSF} and {\tt EXP} types). This amounts to a catalog of approximately 8 million galaxies that will be used in this work. In future studies we plan to use source catalogs produced for UNIONS when available.

\section{Strong Lens Classifier} \label{sec:search}
In recent years, machine learning, and in particular convolutional neural networks (CNNs) have proven to be highly effective for image classification tasks, consistently outperforming ``classical" algorithmic approaches. We adopt the highly successful residual convolutional neural network (ResNet) architecture originally reported in \citet{lanusse2018a} and improved upon in each of \citet{huang2020a} and \citet{huang2021a}
in order to search UNIONS for strong lenses.
In the following sections, we describe the training sample (\S\ref{sec:train}), the ResNet architecture (\S\ref{sec:model}), our training procedure (\S\ref{sec:training}), and our results (\S\ref{sec:training_res}). 
    \subsection{Training Sample}\label{sec:train}

 \citet{huang2020a} described one of the first successful searches for strong lenses that opted to train a ResNet using real images of lenses and ``non-lenses". Here ``non-lens" refers to an image that contains anything but strong lensing (e.g., galaxies, clusters of galaxies, stars, artifacts, or cosmic rays). These non-lenses, specifically a large and diverse sample of them, play a critical role in properly training the ResNet to reject contaminants that it may see in deployment. This strategy has only become more viable with the increased rate of lens discovery in recent years as shown in \citet{huang2021a} and \citet{storfer2024a}. It is clear that the benefit of a limited number of high-quality examples of lenses from ``in-sample" data is equivalent to or better than large, simulated samples. We therefore adopt this strategy in this work. Our training sample consists of 773 images of confirmed and candidate strong lenses. This selection includes systems reported in various imaging  \citep{storfer2024a,shu2022a,stein2021a,savary2021a,huang2020a,huang2021a,canameras2020a,canameras2021a,wong2022a,sonnenfeld2018a,wong2018a,jaelani2020a,sonnenfeld2020a,stark2013a,belokurov2007a,kubo2009a,hennawi2008a,more2012a,sonnenfeld2013a} and spectroscopic searches
 \citep{brownstein2012a,shu2016a,talbot2021a}. These lenses were selected from a parent sample of roughly $1500$ strong lenses at $\delta>30^{\circ}$ in the NGC. We apply cuts with respect to data availability ($gri$ coverage in UNIONS), quality cuts following visual inspection of the systems in UNIONS imaging data, and the $z$-band magnitude cut (see $\S\ref{sec:tractor}$). The breadth of UNIONS enables the construction of a sample of this size while the image quality and depth allows for great diversity in lens characteristics, primarily in their angular sizes and redshifts. 
 
For the non-lens portion of our training sample, we select 36,000 galaxies from \textit{the Tractor} catalog. This amounts to about one lens for every 47 non-lenses in the training sample. These galaxies are selected randomly but approximately proportional to the fractional count of lenses across the survey footprint. This is to prevent the ResNet from learning potential biases. For example, biases may be internal, from inhomogeneity in the UNIONS depth or external, from the use of an out-of-sample object catalog. In order to avoid contamination, we cross-reference this selection with the parent sample of 1500 lenses (i.e., including those not included in the training sample). We also use an early version of our ResNet trained on this sample to remove 78 objects (about half of which are genuine strong lens candidates) from the non-lenses sample via visual inspection of the top 1\% of the ResNet predictions (see $\S$\ref{sec:training} for details on ResNet training). The genuine lenses among these potential contaminants are recovered in ResNet deployment (see $\S$\ref{sec:deploy}). The general expectation is that there is one strong lens in anywhere between $10^3–10^4$ galaxies \citep[e.g.][]{collett2015a}. We do not expect many additional contaminants nor would we expect them to significantly influence the outcome of training.

    \subsection{CNN Architecture: Shielded ResNet Model}\label{sec:model}
    The CNN architecture used in this work was designed based on the residual network (ResNet) architecture described in \citet{he2015a}. 
Within this ResNet architecture, the ``identity" (input) is passed via a ``skip" connection to the output of a convolutional layer or groups of layers \citep[see figure 2 from][]{he2015a}. This allows for information to propagate directly through the portions of architecture linked by the skip connection \citep[][]{he2016a}. Gradients can then be easily calculated during backpropagation in an attempt to avoid the ``vanishing gradient" problem. 
Thus, such ResNet architectures allow for deep CNNs to exceed the previous upper limit on the number of layers (around 6). 
The specific architecture used in this work, is the 46-layer deep \texttt{CMU-Deeplens}, which was originally introduced in \citet{lanusse2018a}.
\citet{huang2020a} improved the performance of the ResNet by re-implementing the architecture in TensorFlow.\footnote{\href{https://www.tensorflow.org/}{https://www.tensorflow.org/}}
\citet{huang2021a} introduced additional changes by including so-called ``shield" layers. This idea was first presented in \citet{szegedy2014a} with the InceptionNet architecture. These shield layers consist of $1\times1$ convolutions that are strategically applied in order to reduce the dimensionality of the output to a desired size. 
This reduces the number of trainable parameters within the ResNet by a factor of 50 and thus increases computational efficiency in terms of memory usage and training time.
Both \citet{huang2021a} and \citet{storfer2024a} show that the addition of shield layers is an improvement over the original ResNet-based architecture from \citet{lanusse2018a} and of the TensorFlow implementation in \citet{huang2020a}. 

    \subsection{Training Procedure}\label{sec:training}
    Prior to training we collect image data for lenses and non-lenses in the training sample that are $140\times140$ pixels ($24^{\prime\prime}\times24^{\prime\prime}$). This cutout size in principle allows for the discovery of lensing systems ranging from galaxy to cluster scales. The training sample is then split into two subsets, the ``training set" and ``validation set", at a 7:3 ratio in order to ensure sufficient data is available to accurately assess model performance. The training set contains 541 and 25078 non-lenses while the validation set contains 232 lenses and 10748 non-lenses. 
Due to the inhomogeneity of the survey between filters we apply a pre-processing routine to the training and validation sets based on statistics from the training set. In particular we use interquartile range (IQR) normalization defined as
\begin{equation}
I_{\text{norm}} = \frac{I - \text{med}(I)}{\text{IQR}(I)},
\end{equation}
where $I$ is the distribution of pixel values and IQR($I$) =  $P_{75}$ - $P_{25}$  (75th -- 25th percentile).
We also clip the pixels values at the 99.9th percentile on the high and low ends of the distributions. This specific pre-processing is in principle robust to not only vastly differing non-Gaussian, pixel-value distributions, but also to extreme outlier values.

 For typical feed-forward networks like the one used here, images are supplied to the ResNet in small equal-sized ``batches". 
 During training the ResNet attempts to minimize the cross-entropy loss function using the training set data:\\
\begin{linenomath*}
\begin{equation}
\label{eqn:loss}
    \displaystyle\mathcal{L}_\text{CE} = -\sum_{i=1}^{N} y_i \log \hat{y}_i+(1-y_i) \log (1-\hat{y}_i)
\end{equation}
\end{linenomath*}
where $y_i$ is the label for the $i$th image (1 for lens and 0 for non-lens), $\hat{y}_i \in [0,1]$ is the model predicted probability, and $N$ is the number of images in each batch. The loss is calculated for each training step, after all images in the batch pass through the ResNet. One ``epoch" of training is completed after all images (or batches) pass through the ResNet once. 

During training, decisions regarding ResNet hyperparameters including the size of the aforementioned batches and number of epochs to train, are critical in optimizing model performance. In this work we choose to show training results from our top two models trained using the same ResNet architecture and training/validation splits with different choices for hyperparameters. We will refer to these two models as Model\,1 (M1) and Model\,2 (M2).
We deem the hyperparameters that are most critical to model performance to be the batch size, initial learning rate, the learning rate decay schedule, learning rate decay factor, and the number of total training epochs.
The initial learning rate is the starting step size for updating model weights, affecting the speed of convergence during model training. The learning rate decay schedule specifies the epochs at which the learning rate is reduced, helping to fine-tune convergence, while the decay factor determines the multiplicative reduction applied to the learning rate in the following epochs. The number of epochs defines the total passes through the full training set. 
For M1 and M2, respectively, we use a batch size of 128 and 64, initial learning rate of 0.002 and 0.001, and a learning rate decay at the epochs of [60, 100] and [40, 80]. For both models we use a learning rate decay factor of [0.5,0.1] and train for a total of 120 epochs.

Implementation of this ResNet architecture in \texttt{TensorFlow} has allowed for relatively simple implementation of distributed training across multiple graphics processing units (GPUs) using \texttt{Tensorflow}'s MirroredStrategy. MirroredStrategy enables synchronous data-parallel training across multiple GPUs on a single machine by mirroring model variables across devices. It splits the input batch evenly among GPUs, allowing each device to compute forward and backward passes independently. Gradients are aggregated using an all-reduce operation, ensuring synchronized updates to model weights. This strategy is efficient and scalable, decreasing training time approximately linearly based on the number of GPUs used. In this work, training is done across four NVIDIA V100 GPUs on a single node, made available by the University of Hawaii's high-performance computing (HPC) cluster, Koa. The batch size for each model refers to the size of the batch distributed to each GPU. Total training time is extremely fast at only $40$ seconds per epoch or $80$ minutes total.

We choose to evaluate the models using the receiver operating characteristic (ROC) curve and the Precision versus Recall Curve (PRC). The ROC curve shows the True Positive Rate (TPR) versus the False Positive Rate (FPR) which are defined as
\begin{equation}
    \rm{TPR} = \frac{N_{\rm{TP}}}{N_{\rm{P}}} = \frac{N_{\rm{TP}}}{N_{\rm{TP}} + N_{\rm{FN}}}
\end{equation}

and
\begin{equation}
    \rm{FPR} = \frac{N_{\rm{FP}}}{N_{\rm{N}}} = \frac{N_{\rm{FP}}}{N_{\rm{FP}} + N_{\rm{TN}}},
\end{equation}
where positives (N$_{\rm{P}}$) indicates the total number of lenses, negatives (N$_{\rm{N}}$) indicates the total number of non-lenses, True Positives (N$_{\rm{TP}}$) is the number of correctly classified lenses, False Positives (N$_{\rm{FP}}$) is the number of non-lenses incorrectly classified as lenses, and False Negatives (N$_{\rm{FN}}$) is the number of lenses incorrectly classified as non-lenses. The curve is generated by varying the probability threshold for classification. The area under the ROC curve (or ``AUC") is perhaps the best single number metric to judge model performance. An AUC = 0.5 indicates a random classifier, and AUC = 1 a perfect classifier. 

In the case of the PRC, precision is defined as:
\begin{equation}
    \rm{Precision} = \frac{N_{\rm{TP}}}{N_{\rm{TP}} + N_{\rm{FP}}}
\end{equation}
Precision quantifies the rate of correct positive classifications among all positive classifications whereas the recall is exactly equivalent to the TPR. While classically, the ROC curve is the most prevalent, the PRC is particularly valuable when considering datasets with high class imbalance as is the case here, in the training sample (1:46, lens:non-lens ratio) and in eventual deployment (1 lens for every $\mathcal{O}(10^4)$ galaxies). Precision directly captures the impact of false positives which is especially crucial when the cost of false positives is high.

    \subsection{Training Results}\label{sec:training_res}
    To assess model performance during training we evaluate the AUC and cross-entropy loss at every training step. Figure~\ref{fig:auc-loss} shows the AUC (left) and the loss (right) at each epoch for both the training and validation sets for M1 and M2. During training the AUC is monitored and the model is saved at the epoch with the highest AUC (indicated by the stars), as well as at the completion of training (epoch 120). The highest AUC for M1 is achieved at epoch 102 and for M2 at 110. The early saving mechanism ensures that the best performing epoch is retained. This early saving is seen to occur prior to any signs of dramatic divergence in training performance metrics in both models, more easily discernible in the loss curves. While both models perform similarly well, each displays a clear generalization gap. This is the difference between the model's performance on the training set compared with the validation set, indicating how well the model generalizes to new inputs. This gap is a clear indication that the models have not converged and may be indicative of a shortcoming in the training sample, training hyperparameters, or data pre-processing. However, we opt to proceed with these trained models, since they perform best after a comparison of standard model performance metrics, including those described in $\S$~\ref{sec:training}. 

\begin{minipage}{0.99\textwidth}
  \vspace{0.2cm}
\makebox[\textwidth][c]{
  \includegraphics[keepaspectratio=true,width=0.99\textwidth]{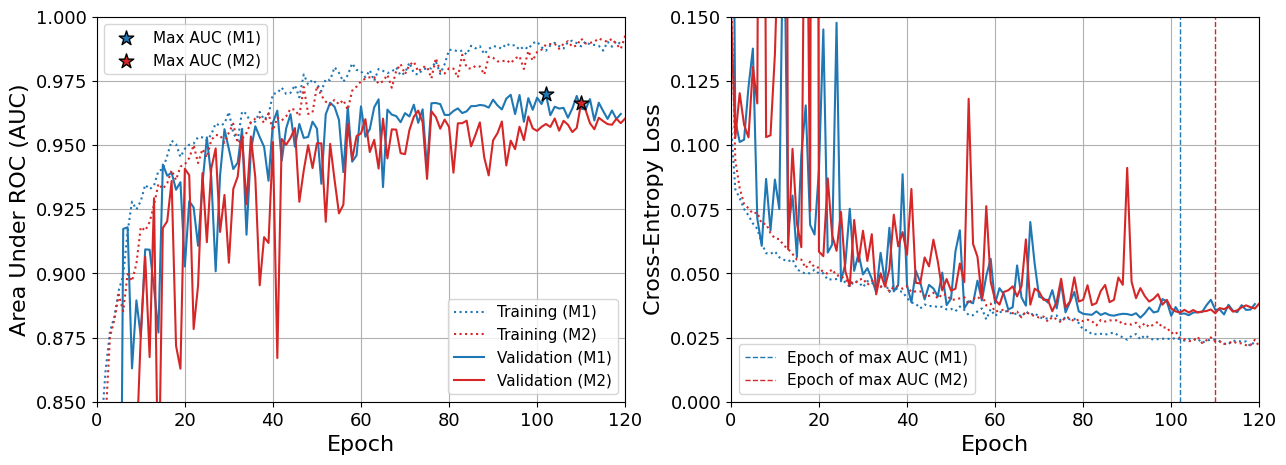}}
\captionof{figure}{\footnotesize AUC (left) and cross-entropy loss (right) for both training and validation sets over the course of the ResNet training for both models used in this work. Each panel shows the metrics for the validation set in solid lines and the training set in dotted lines, with M1 shown in blue and M2 shown in red. The stars in the left panel indicate the location and value of the maximum AUC for each model during training. Correspondingly, the right panel shows vertical dashed lines to indicate the epoch of the maximum AUC.}
\label{fig:auc-loss}
\end{minipage}
\vspace{0.5cm}

The ROC curve and PRC are shown in Figure~\ref{fig:roc} (left and right panels respectively), evaluated on the validation set at the best saved model epoch. The curves for M1 and M2 are shown in blue and red. Note that the FPR in the ROC plot is on a log-scale to highlight small differences in the models in the low FPR regime. While M1 performs slightly better in the AUC (around 1\% overall) we find marginally better performance in both ROC and PRC by simply taking the maximum probability ($\text{max}$(M1,M2)) from either model for each image. The ROC curve produced by this combination of the models is shown in green. On these green curves we also plot the probability thresholds in black. In particular, we see that at most probability thresholds between 0.9 and 0.1 there is a marked improvement in the $\text{max}$(M1,M2) TPR for the same FPR, meaning that we can increase our completeness while maintaining the same level of contamination. The PRC for $\text{max}$(M1,M2) similarly shows a higher recall at the same precision compared with M1 and M2 individually. We discuss the implications for these training results further with respect to our ResNet deployment in $\S$\ref{sec:deploy} and in a broader context in $\S$\ref{sec:discussion}.

\vspace{-0.15cm}
\begin{minipage}{0.99\textwidth}
\makebox[\textwidth][c]{
  \includegraphics[keepaspectratio=true,width=0.92\textwidth]{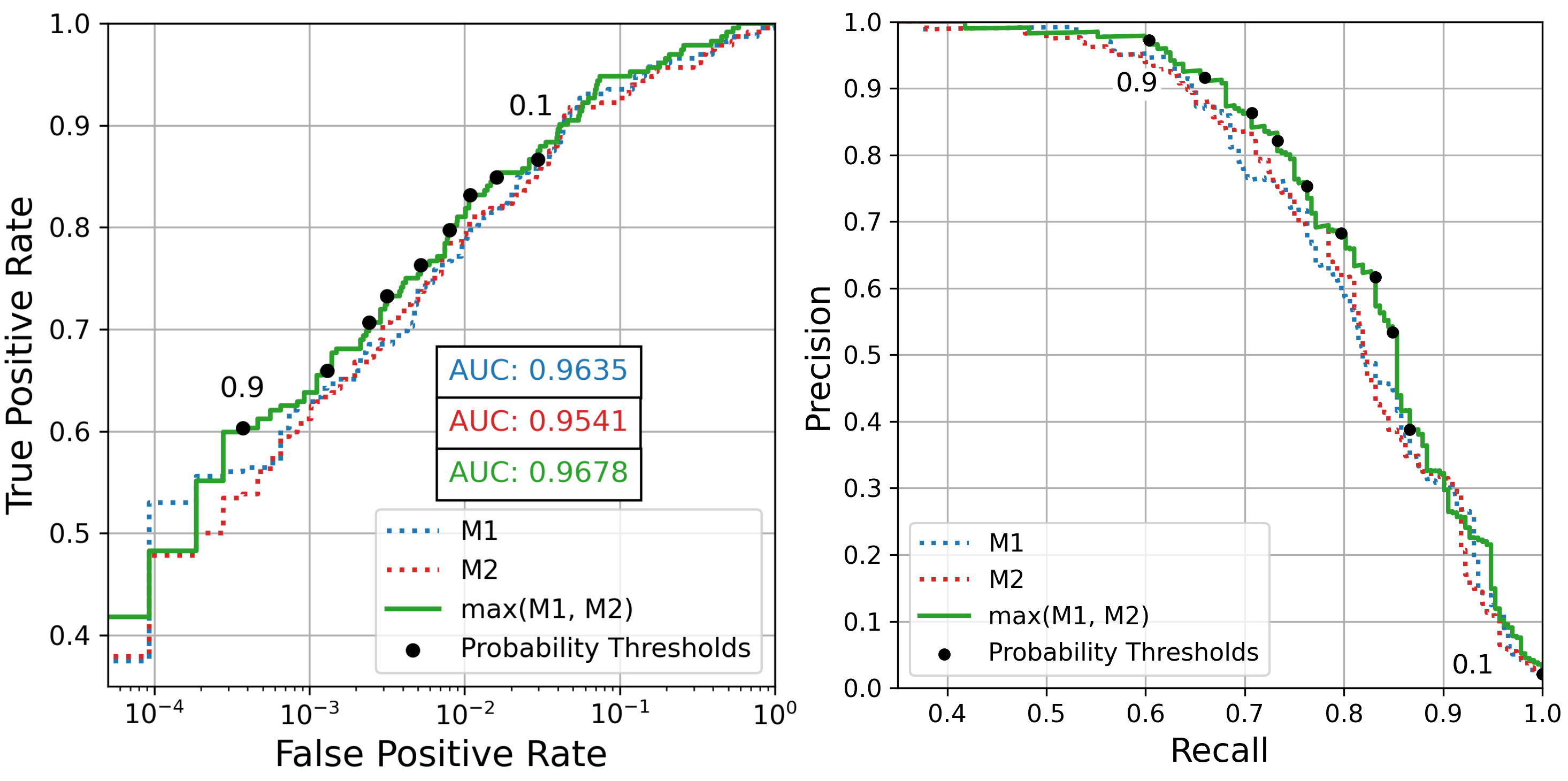}}
\captionof{figure}{\footnotesize TPR versus FPR (ROC, left) and Precision versus Recall (right) for both models used in this work, as well as for their combined performance (maximum prediction). On both panels individual models are shown as dotted lines, with M1 shown in blue, M2 shown in red and their combination shown as a solid green line. Black points indicate probability thresholds for the combination of the models for each plot. Note that the left panel shows the range (0.35, 1.0) with the $x$-axis on a log-scale beginning at 10$^{-4}$.}
\label{fig:roc}
\end{minipage}


\section{New Strong Lens Candidates} \label{sec:results}
In this section we describe the deployment of our ResNet models to UNIONS ($\S$~\ref{sec:deploy}), outline visual inspection of the ResNet recommendations ($\S$~\ref{sec:inspection}), and report our strong lens candidates ($\S$~\ref{sec:candidates}). 

\subsection{Deployment}
\label{sec:deploy}
Our deployment sample consists of the 8 million galaxies in the DESI LS catalog from \textit{the Tractor}. We collect 140$\times$140 pixel cutout images in the $gri$ bands, centered on each object contained within a skycell (0.5$^{\circ}$$\times$0.5$^{\circ}$). All cutout images from each skycell are then scaled using the IQR and clipped at the 99.9th percentile based on statistics from the training sample, similar to that of the validation set, as described in $\S$~\ref{sec:training}. We conduct our deployment following an on-the-fly strategy where all target galaxies in each skycell are processed together and ingested by our two ResNet models, in real time. This alleviates data storage requirements by only retaining data from the full skycell in memory and writing desired outputs, which also saves computational time from I/O operations. 

Prior to deployment we must set a predetermined threshold on the ResNet predicted probabilities. We decide to use a percentile threshold cut, since the primary limitation for visual inspection of the ResNet recommendations is human-time. 
We determine this threshold by taking the probability corresponding to the 99.9th percentile of all probabilities for images in deployment from each model individually. This corresponds to thresholds of 0.8712 and 0.8796 for M1 and M2 respectively. 
In practice, we apply this threshold cut to the deployment sample so that an image must receive a predicted probability that exceeds the respective threshold for either or both models. Algorithmically this is defined simply as
\begin{equation}
    (P_\text{M1}(x_n) > t_\text{M1}) \lor (P_\text{M2}(x_n) > t_\text{M2}) \equiv \text{max}(P_\text{M1}(x_n),P_\text{M1}(x_n)) > t_{\text{eff}}    
\end{equation}
where $x_n$ is the $gri$ image data for object $n$, $P_\text{M1}$ and $P_\text{M2}$ is the predicted probability from M1 and M2, $t_\text{M1}$ and $t_\text{M2}$ are their respective probability thresholds, and $t_{\text{eff}}$ is some effective probability threshold. Given our cuts, $t_{\text{eff}} = t_\text{M1} = 0.8712$.

Following the application of probability threshold cuts, we are left with a sample of $11600$ images which we refer to as ``recommendations" from the ResNet. We then removed ``duplicate" images which are cutouts centered on different objects that overlap significantly. Known strong lensing candidates are also removed, including systems which were not used in the training sample. Following this, we are left with $10400$ recommendations, all subject to visual inspection.
\subsection{Visual Inspection}
\label{sec:inspection}
Visual inspection (VI) of the recommendation images was conducted by co-authors C.J.~Storfer and E.~Magnier. When identifying strong lens candidates, we expect most often to find small blue galaxies near a massive galaxy, usually with elliptical profiles (Sérsic index, n$_\text{s}>2$). These blue galaxies are the most frequently detected type of lensed galaxy (though we sometimes find red galaxies as sources as well). Similar to the characteristics described in \citet{huang2020a}, \citet{huang2021a}, and \citet{storfer2024a}, these lensed sources typically:
\begin{enumerate}
    \item are 0.5–5$^{\prime\prime}$ away from the lens galaxy/galaxies;
    \item are tangentially elongated;
    \item curve toward the lens galaxy/galaxies (including semi- or nearly full rings);
    \item have counter/multiple images with similar colors;
    \item have a relatively low surface brightness.
\end{enumerate}
These characteristics are heavily skewed towards the most common type of lensing scenario, galaxy-galaxy lensing. In many of these cases, systems can be identified as variations of ``archetypal" lensing configurations \citep[see][Figure A1]{gu2022a}.
Typically, most candidates do not have all the above characteristics. In general, the greater the number of these characteristics
an image has, the higher they are ranked in VI. However, in group- or cluster-scale lensing scenarios there is a wide variation in characteristics, which may be highly non-standard. This includes very large separations between lens and source images and highly magnified and/or irregular source image morphologies. Due to the low numbers of expected group- and cluster-scale lenses we primarily adhere to the above listed criteria, although with some exceptions.

For VI, four-panel images including the \textit{gri} color image and each of the individual filter images were created for all recommendations. An example of this is shown in Figure~\ref{fig:inspection-panel}. In this process, the individual $g$ and $r$-bands were extremely helpful in identifying lensed images/arcs. This is due to their exquisite image quality and for the $g$-band, high contrast in lens versus source colors. This revealed a number of counter-images which may have otherwise been missed due to lens light contamination. The initial VI of the $10400$ recommendations was conducted by author C.J.S., resulting in the removal of about $7500$ false-positives. The remaining $2500$ recommendations were then assigned an initial letter grade by C.J.S. The grading criteria were similar to those in \citet{storfer2024a} and are described as follows.
\begin{itemize}
    \item \textit{Grade A:} We have a high level of confidence of these candidates. Many of them have one or more prominent arcs, usually blue. The rest have one or more clear arclets, sometimes arranged in multiple-image configurations (e.g. an Einstein cross) with similar colors (again, typically blue). However, there are some clear cases with red arcs. Most of these systems can be easily classified as one of the ``archetypal" lensing configurations.
    \item \textit{Grade B:} They have similar characteristics as the Grade As. However, the lensed arcs tend to be fainter/lower signal-to-noise than those for Grade A. Likewise, the putative arclets tend to be smaller and/or fainter, or isolated (without counterimages). Occasionally these systems may exhibit characteristics that do not follow the ``archetypal" configurations.
    \item \textit{Grade C:} They generally have features that are even fainter and/or smaller than what is typical for Grade B candidates, but that are nevertheless suggestive of lensed arclets. Counterimages are often not present or hardly discernible. In a number of cases, the angular scales of the candidate systems are comparable to or only slightly larger than the seeing. Additional evidence or data may be required. 
\end{itemize}

\begin{minipage}{0.99\textwidth}
  \vspace{0cm}
\makebox[\textwidth][c]{
  \includegraphics[keepaspectratio=true,width=0.95\textwidth]{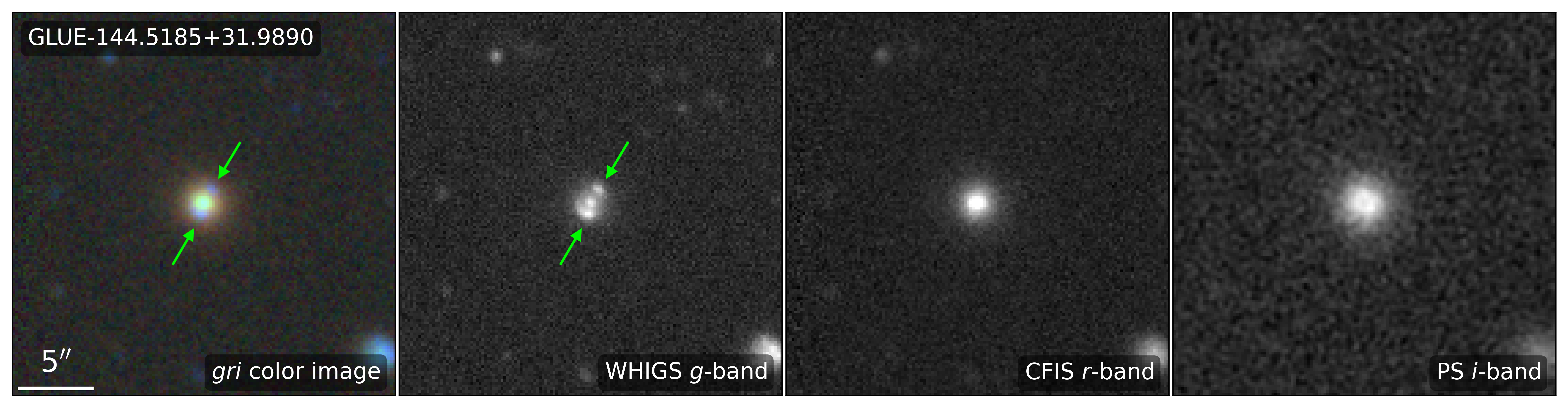}}
\captionof{figure}{\footnotesize Example of an image panel used in visual inspection, displaying the color image and individual filter images for a new strong lens candidate reported here. The system name is displayed in the top left of the first image panel with a 5$^{\prime\prime}$ scale-bar shown at the bottom left. The filter or filter combination is shown in the bottom right of each panel. Green arrows were added to the color and the $g$-band image to indicate the position of the lensed images. These labels and arrows were not present during inspection and were added later.}
\label{fig:inspection-panel}
\vspace{0.2cm}
\end{minipage}

It is important to note that grades were assigned based on the (subjective) relative likelihood of an image containing strong lensing, not the potential scientific quality although there is a strong correlation. 
For example, given additional evidence in the form of spectroscopy or high-resolution imaging, many of the grade C systems may be confirmed to be true lenses and may prove to be useful for various cosmology and galaxy-evolution studies. We grade strictly on the quality of evidence provided from the UNIONS \textit{gri} imaging data. All systems receiving a grade of C or above are regarded as candidates. Note that we maintain a high standard even for grade C systems. 

\subsection{New Strong Lens Candidates}
\label{sec:candidates}
\begin{figure}[ht!]
  \includegraphics[keepaspectratio=true,width=0.95\textwidth]{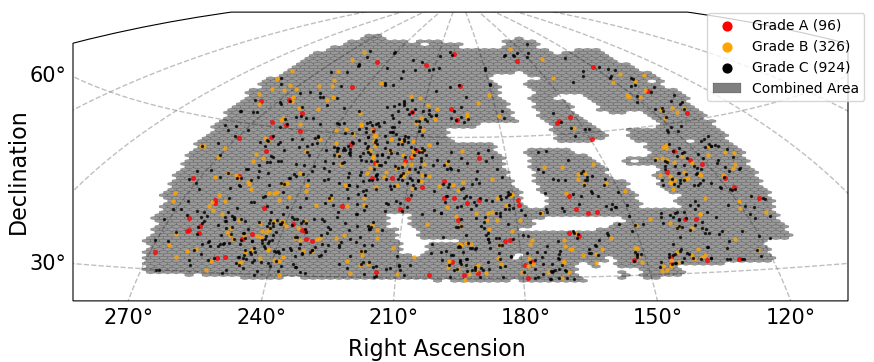}
  \vspace{-0.25cm}
\caption{\footnotesize Map indicating the location of \totalprelim new candidates over-plotted on the combined UNIONS $gri$ area.}
\label{fig:cand-map}
\end{figure}

Following the initial round of inspection and grading, C.J.S. found a total of \totalprelim new strong lens candidates of which \gradeAprelim are grade A, \gradeBprelim grade B, and \gradeCprelim grade C. These candidates were then re-inspected by co-author E.A.M. Letter grades were converted to numerical scores and averaged with A (4), B (3), and C (2). We consider candidates with an average score of 3.5 or 4 to be grade A, 3 to be grade B, and 2 or 2.5 to be grade C. In total we find \gradeAfinal grade As (4s: 62, 3.5s: 84), \gradeBfinal grade Bs (3s: 199), and \gradeCfinal grade Cs (2.5s: 412, 2s: 589). 
We report the average score for each candidate for transparency. 
The locations of the candidates in the UNIONS footprint are shown in Figure~\ref{fig:cand-map} and a representative sample of candidates from each grade is shown in Figure~\ref{fig:cand-examples}. We also make the full catalog of candidates, along with image cutouts, publicly available on Zendodo.\footnote{\href{https://zenodo.org/records/15355369}{Link to full catalog and candidates}}

\vspace{0.5cm}
\vspace{-0.5cm}

\begin{figure}[ht!]
    \centering
    \includegraphics[width=0.98\textwidth]{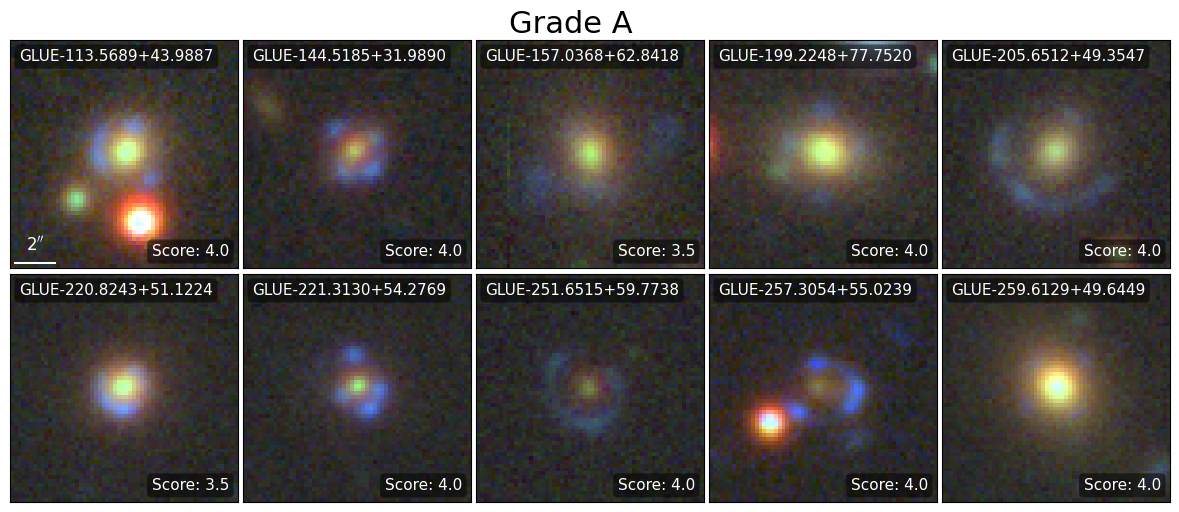}

    \caption{Representative sample of new candidates reported in this work. Each image panel has N up and E left, and system names are displayed in the upper left corner of each image panel. The average score from inspectors C.S. and E.A.M. are shown in the bottom right corner of each image panel. Rows 1 and 2 show grade A candidates, rows 3 and 4 show grade B candidates, and rows 5–7 show grade C candidates.}
    \label{fig:cand-examples}
\end{figure}
\clearpage
\begin{figure}[ht!]
    \figurenum{\ref{fig:cand-examples}} 
    \centering
    \includegraphics[width=0.98\textwidth]{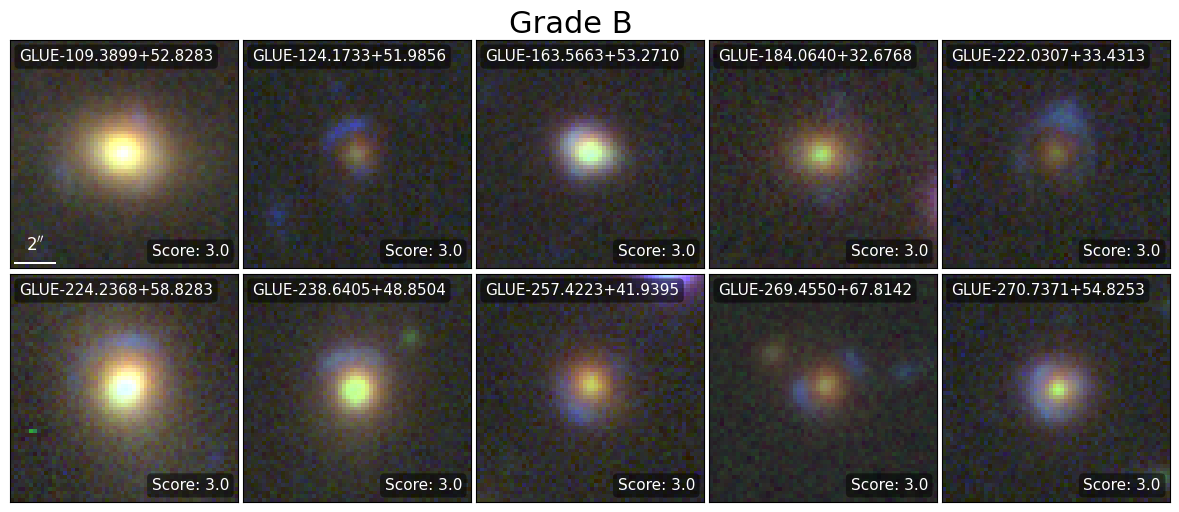}
    \includegraphics[width=0.98\textwidth]{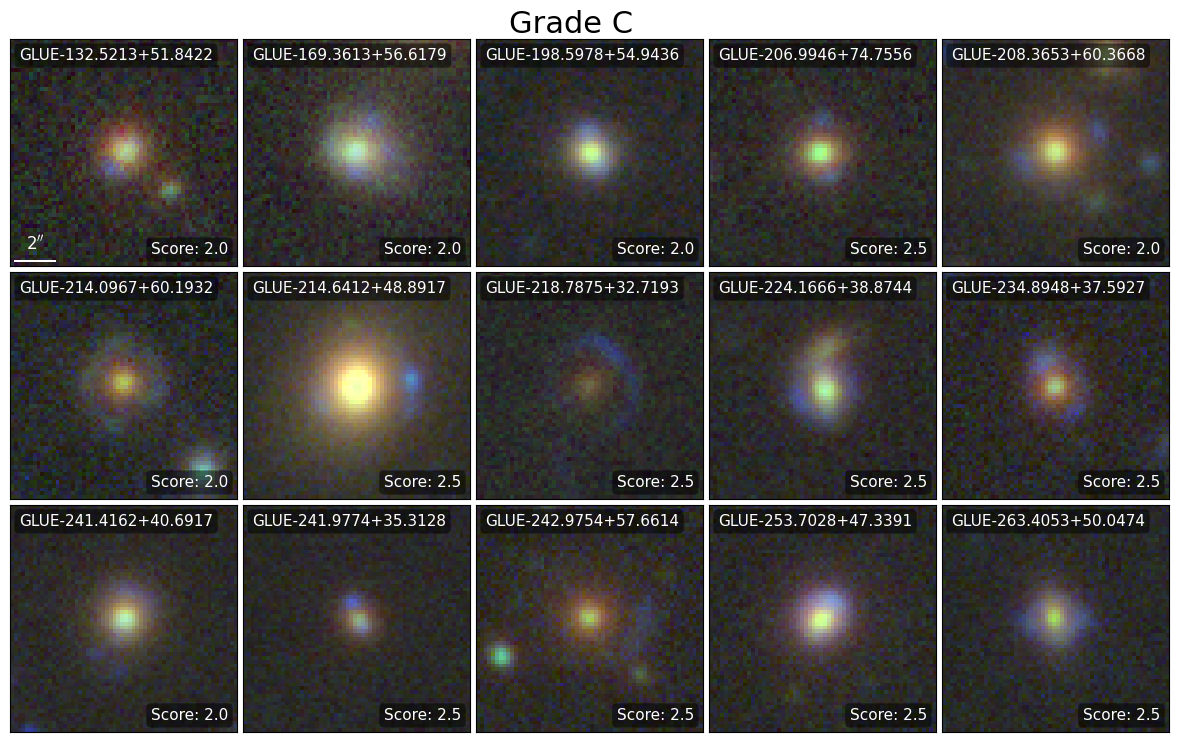}
    \caption{(Continued.)}
\end{figure}

\section{Spectroscopy of Strong Lens Candidates} \label{sec:spectroscopy}
\subsection{Spectroscopic Data}
\label{sec:spectro}
\begin{figure}[ht!]
    \centering
    \makebox[\textwidth]{%
        \includegraphics[width=0.514\textwidth]{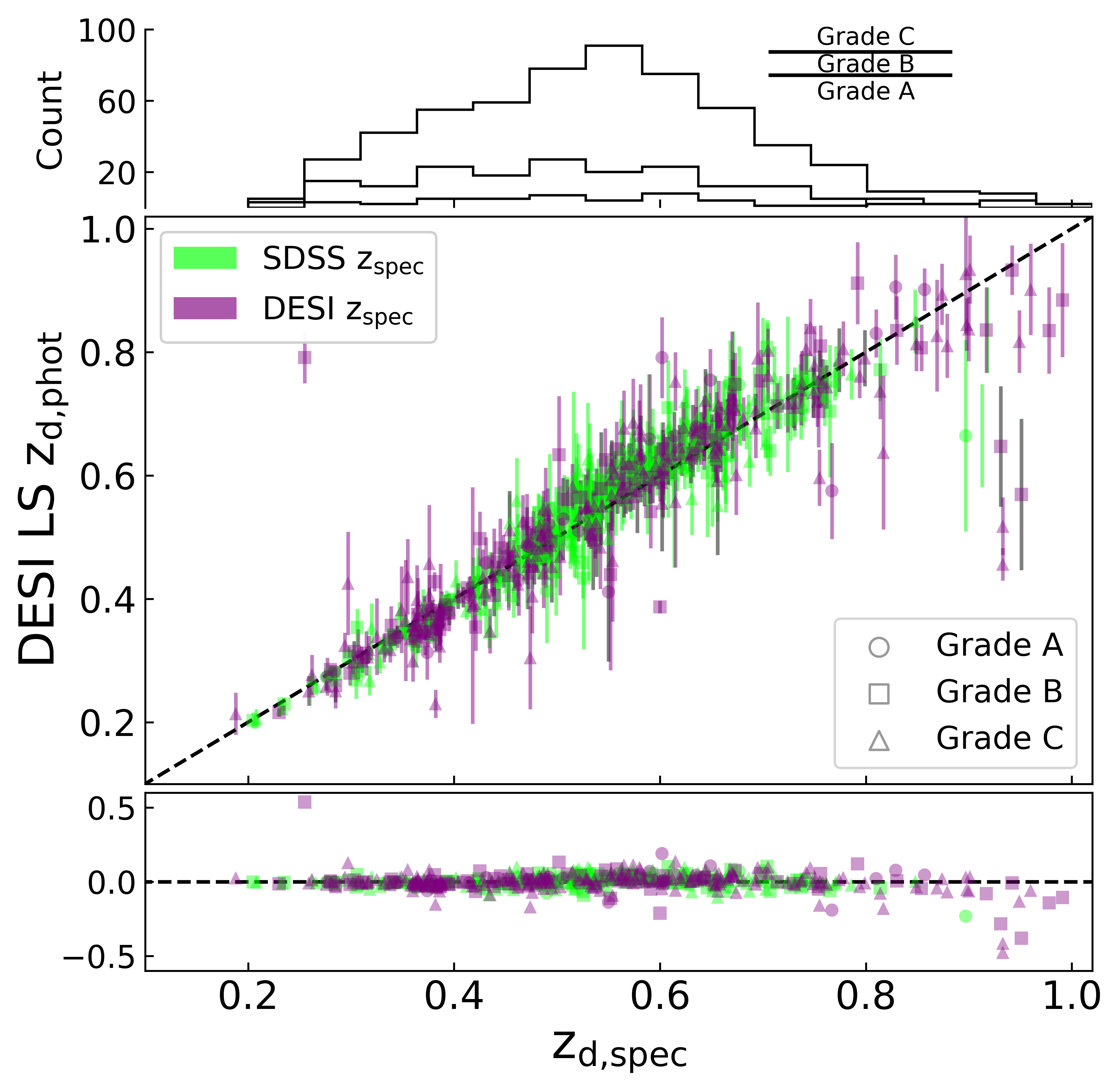}
        \hspace{-0mm}
        \includegraphics[width=0.45\textwidth]{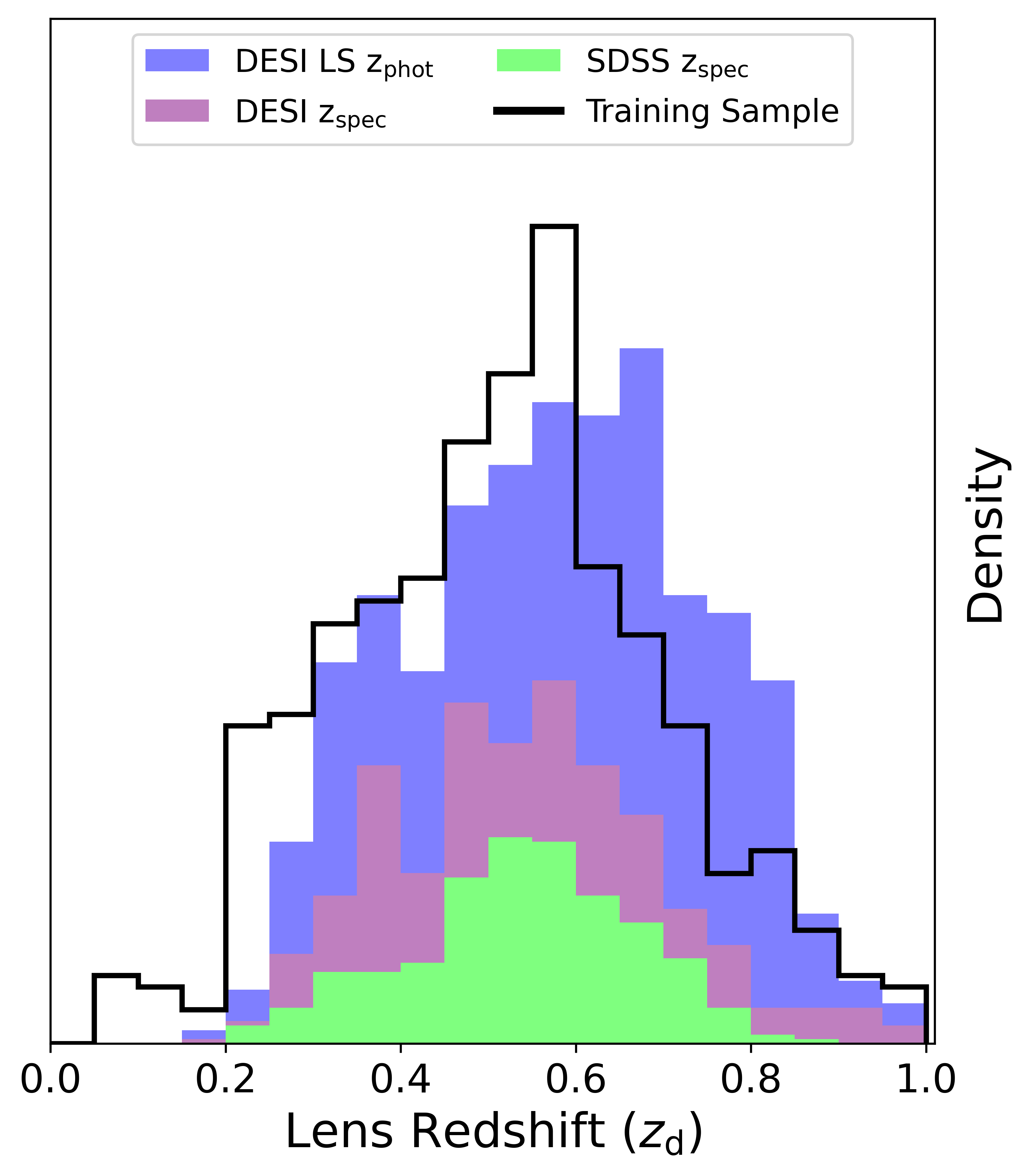}
         \vspace{-1cm}
    }
   
    \caption{\footnotesize \textit{Left:} DESI LS photometric redshifts plotted against either SDSS (eBOSS) DR18 (green) or DESI EDR (purple) spectroscopic redshifts where available for lens galaxies. The marker type indicates the visual inspection grade. The top panel shows an accompanying stacked histogram of the lens redshift distribution separated by visual inspection grade (bottom to top: A, B, C). The bottom panel shows the residual of spectroscopic and photometric redshifts. \textit{Right:} Distribution of $z_\text{d}$ (spectroscopic and photometric) for all candidate lenses plotted as color-filled, stacked histograms. Green, purple, and blue colors indicate the source of the redshift measurement from either SDSS (eBOSS), DESI, or DESI LS. The black histogram shows the combined distribution of $z_\text{d}$ for the lenses in the training sample with redshift data from the same sources listed above (SDSS, DESI, or DESI LS). Both histograms are normalized to have the same probability density.}
    \label{fig:spec-phot}
\end{figure}

\begin{figure}[H]
    \centering
    \includegraphics[width=0.99\textwidth]{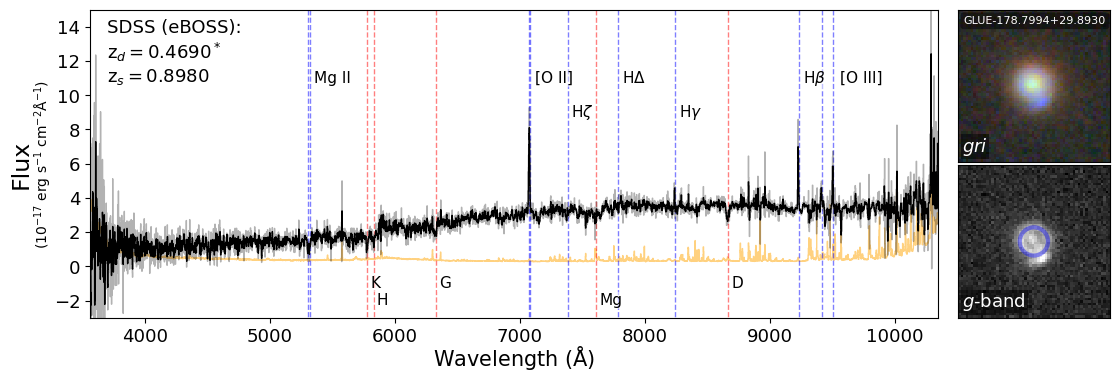}
    \includegraphics[width=0.99\textwidth]{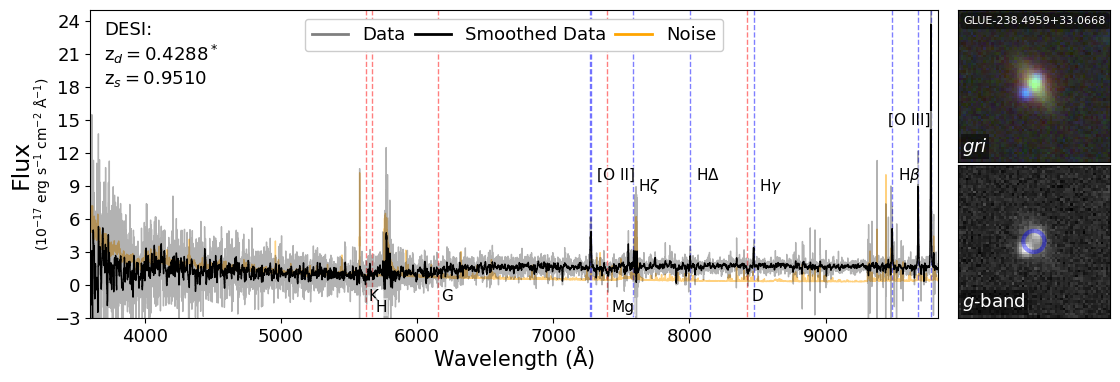}
    \includegraphics[width=0.99\textwidth]{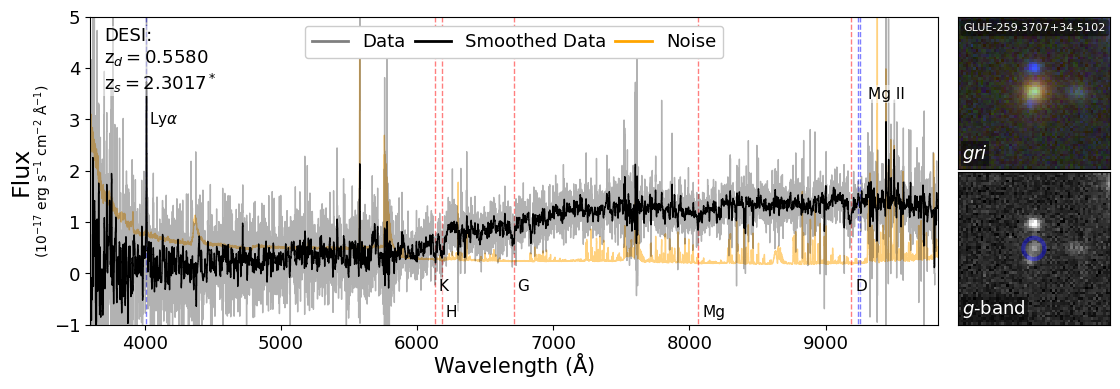}
    \caption{\footnotesize Three of the fully confirmed systems from the sample of candidates reported in this work. The data show strong evidence for superposition of spectra for two galaxies at different redshifts. For each of the panels, we show the spectrum from either DESI or SDSS. 
    Plotted are the original spectrum (gray), the Gaussian smoothed spectrum (black), and the noise spectrum (orange). We show the lens (red) and source (blue) spectral features with labels. The source of the data (DESI or SDSS) is shown in the top left of this panel along with lens (z$_\text{d}$) and source (z$_\text{s}$) redshifts (an asterisk indicates manual redshift fits). 
    The UNIONS $gri$ image is shown to the upper right with the WHIGS $g$-band image below that. The respective survey fiber size and location are indicated by the blue circle in the $g$-band image. 
    The three panels show GLUE-178.7994+29.8930, GLUE-238.4959+33.0668, and GLUE-259.3707+34.5102 which received grades of A, B, and B respectively, confirmed via SDSS and/or DESI spectroscopy.}
    \label{fig:example}
\end{figure}

Of the \totalfinal new lenses reported in this work, 
580 lens galaxies have spectroscopic redshifts ($z_\text{d}$): 
283 come from the Sloan Digital Sky Survey (SDSS) Extended Baryon Oscillation Spectroscopic Survey (eBOSS) Data Release 18 (DR18); and 297 come from the DESI Data Release 1 (DR1). Of the 297 systems with DESI spectra, 129 also have SDSS spectra. We opt to use the DESI data in these instances for the analysis which follows.
We compare the spectroscopic and photometric redshifts from the DESI LS \citep{zhou2020a} for each of these objects shown in Figure~\ref{fig:spec-phot},
along with the distribution of $z_\text{d}$ for all candidates (and the training sample). 
We find that the photometric redshifts are generally excellent and in very close agreement with their ground-truth spectroscopic redshifts with the average difference of $-$3.9$\times10^{-3}$ with a standard deviation of 8.2$\times10^{-2}$. 
There are, however, some dramatic outliers.
A total of 49 candidates have a difference between the measured spectroscopic and photometric redshift $>1\sigma$ (8.2$\times10^{-2}$). We inspect the quality of the photometry and also of the redshift fit in all instances. We find that all of these systems have reliable photometry, but find 15 instances where the spectrum displays a evidence of a superposition of spectral features at different redshifts (three from SDSS and 12 from DESI). We show three examples of these systems in Figure~\ref{fig:example} with the full sample shown in Table~\ref{tab:spec-conf}. The table includes a quality score for our confidence in the source redshift exclusively, where ``1" is definite and ``2" is probable. This scoring is based on the number of background emission features present, and their signal-to-noise ratio. In a majority of these instances, background source emission is high signal-to-noise causing the default redshift-fitting pipelines for SDSS \citep{bolton2012a} and DESI (Redrock\footnote{\href{https://github.com/desihub/redrock}{https://github.com/desihub/redrock}}; S.~J.~Bailey et al., in prep.) to preferentially fit to spectral features of the background source galaxy. We were able to also identify absorption features typical of early-type, elliptical lens galaxies at redshifts more similar to the original photometric redshift. 
These systems are thus all fully spectroscopically confirmed. Many of the remaining 34 systems display some abnormalities in the spectra and in the spectroscopic redshift fit but do not contain sufficient evidence for confirmation. However, we do not rule out these systems as candidates, despite the lack of evidence for source redshift features. 

This methodology -- identifying a discrepancy between spectroscopic and photometric redshifts -- could be a promising approach for discovering new spectroscopic lenses.
This would be distinct from past spectroscopic searches, such as those of \citet{bolton2006a}, \citet{brownstein2012a}, and \citet{talbot2021a}, which identify emission-line source galaxies in spectra of photometrically identified luminous red galaxies (LRGs) by subtracting a model of the LRG spectrum. In this case, one could search spectra with a large discrepancy between photometric and spectroscopic redshifts regardless of photometric typing. In all of the confirmed systems, including those shown in Figure~\ref{fig:example}, the red, lens-galaxy continuum (in particular the 4000\AA~break) is quite prevalent and source emission is usually high S/N, making these systems easily identifiable, even by eye. Such a search would likely preferentially select highly star-forming source galaxies because it relies, in most instances, on the redshfit being fit to the source galaxy emission features. Candidates discovered this way could then be inspected in UNIONS imaging data for additional evidence. Combining this methodology with the ongoing Euclid space telescope survey or the future Roman space telescope imaging data will provide more definitive evidence of the nature of these systems. 



The total $z_\text{d}$ distribution for the candidates compared with the training sample is shown in the right panel of Figure~\ref{fig:spec-phot}. These distributions are visually slightly different which is reflected in the median redshifts for the candidates and the lenses in the training sample: 0.58 and 0.51, respectively. While this is only a marginal difference, it is clear that the ResNet is capable of finding new lenses at redshifts that may be underrepresented in the training sample. This is especially evident in the redshift 0.60--0.85 bins. 

\subsubsection{Auxiliary Spectroscopic Data}
\label{sec:aux-spec}
In addition to the spectroscopic data utilized from existing spectroscopic surveys, we have also begun independent spectroscopic follow up of the lensed sources from the candidates reported in this work. In particular we target systems with moderate-to-high lens redshifts ($0.6~<~z_\text{d}~<~1.0$; spectroscopic or photometric) with near-infrared (NIR) spectrographs GNIRS on Gemini North and NIRES on Keck II. Based on results from \citet[][]{tran2022a}, Agarwal et al. (in prep.), and Storfer et al. (in prep.), NIR spectroscopy is extraordinarily successful in capturing spectral features of lensed emission-line galaxies at high redshifts ($z_\text{s}\gtrsim$~1.7). Spectroscopic confirmation and redshift measurements of such systems ($z_\text{d}~>~0.6$; $z_\text{s}~>~2.0$) will be of critical importance, in particular in the pursuit of measuring the mass-density profile slope in elliptical galaxies. So far, seven sources for the candidates reported in this work have been observed (two with GNIRS, five with NIRES). All seven have been confirmed with all source redshifts $z_\text{s}$ $>$~2.1. These results will be reported in an upcoming publication. 

\begin{table}[h]
    \vspace{0.25cm}
    \hspace{-2cm}
    \centering
    \begin{tabular}{lcccccc}
   
        \toprule
        \textbf{Name} & \textbf{Grade} & \textbf{Score} & \textbf{$z_\text{d}$} & \textbf{$z_\text{s}$} & \textbf{Survey} & \textbf{Quality}\\
        \midrule
        GLUE-115.7537+67.9583 & C & 2.5 & 0.4351 & 2.8851$^{*}$ & DESI & 2\\
        GLUE-127.8136+32.4168 & A & 4 & 0.4859$^{*}$  & 0.7670 & DESI & 1\\
        GLUE-178.7994+29.8930 & B & 3 & 0.4690$^{*}$   & 0.8980 & SDSS & 1 \\
        GLUE-184.9193+32.9334 & B & 3 & 0.4922$^{*}$  & 0.9310 & DESI & 1\\
        GLUE-185.2226+37.0239 & C & 2.5 & 0.6147 & 1.0000$^{*}$ & DESI & 2\\
        GLUE-193.0230+45.9556 & A & 3.5 & 0.6020 & 2.3630$^{*}$ & DESI & 1\\
        GLUE-198.3077+45.2629 & B & 3 & 0.7199$^{*}$  & 1.2102$^{*}$ & DESI & 1\\
        GLUE-229.9844+38.7616 & C & 2 & 0.5800$^{*}$   & 0.9130 & SDSS & 2 \\
        GLUE-233.6642+50.4589 & C & 2 & 0.7390$^{*}$   & 1.1690 & SDSS & 2 \\
        GLUE-236.3518+39.9062 & B & 3 & 0.6368$^{*}$   & 1.6266 & DESI & 1\\
        GLUE-238.4959+33.0668 & C & 2.5 & 0.4288$^{*}$ & 0.9510 & DESI & 1\\
        GLUE-244.5015+55.6958 & C & 2 & 0.5650   & 1.1020$^{*}$ & DESI & 1\\
        GLUE-259.3707+34.5102 & B & 3 & 0.5580   & 2.3017$^{*}$ & DESI & 1\\
        GLUE-265.5754+65.2324 & C & 2 & 0.5914$^{*}$   & 0.7550 & DESI & 1\\
        GLUE-278.9295+61.9683 & C & 2 & 0.4940$^{*}$   & 0.8167 & DESI & 1\\
        \bottomrule
    \end{tabular}
    \caption{\footnotesize Fully confirmed strong lensing systems in spectroscopic fibers from SDSS DR18 and DESI DR1. Asterisks indicate redshifts fit for manually.}
    \label{tab:spec-conf}
\end{table}
\vspace{-1cm}

\section{Discussion} \label{sec:discussion}
In this section we discuss the overall performance of our ResNet deployment in $\S$\ref{sec:deploy_perf}. We then perform a brief comparison with recent strong lens searches including \citet[][]{barroso2025a} ($\S$~\ref{sec:AB25-comparison}), \citet[][]{storfer2024a} ($\S$\ref{sec:S24-comparison}), and with \citet[][Euclid Strong Lens Discovery Engine]{SLDE-A2025a} ($\S$\ref{sec:SLDE-comparison}). We also discuss implications for the future of UNIONS and strong lens searches with respect to Euclid ($\S$\ref{sec:euclid}).

\subsection{Deployment Performance}
\label{sec:deploy_perf}
In this work, we have trained two ResNet models on real images of lenses and non-lenses from UNIONS. We adopt a probability threshold (0.8712) that corresponds to the top 99.9\% of probabilities for images from either model. Together, the ResNets uncover \totalfinal new strong lens candidates of which we classify \gradeAfinal as Grade A, \gradeBfinal as Grade B, and \gradeCfinal as Grade C. Once again, we maintain a high-standard in visual inspection and thus have high confidence in the Grade C candidates overall.

\begin{figure}[H]
    \centering
    \includegraphics[width=0.98\textwidth]{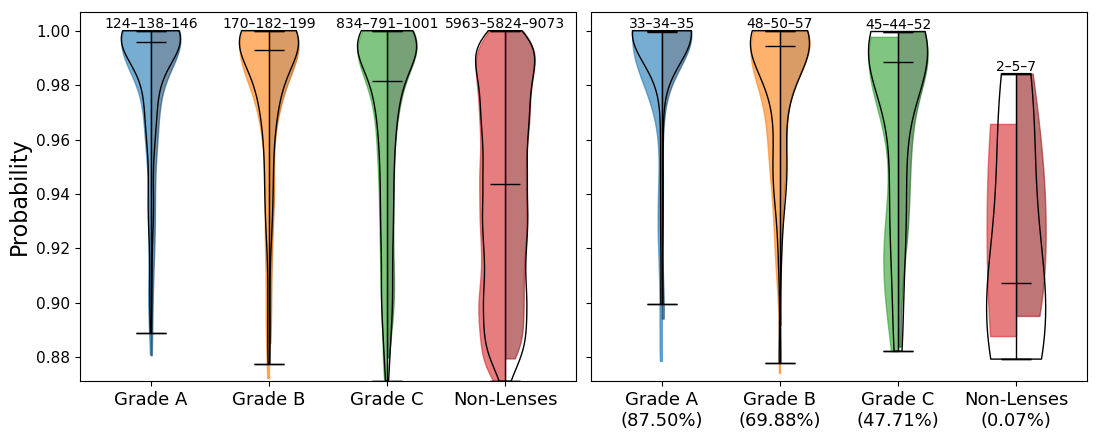}
    \caption{\footnotesize Violin plots for the distribution of ResNet predicted probabilities for images in deployment (left panel) and the validation set (right panel), separated by grade including non-lenses. Each violin shows the distribution for M1 (left) and M2 (right) probabilities above their respective 99.9th percentile probability cut. The black outline shows the maximum probability between M1 and M2 for each image above the effective probability threshold with the horizontal bars on each showing the maximum, median, and minimum probability (top to bottom). The thresholds used in deployment for each model and the combined threshold are applied in the same way, after-the-fact to the validation set (right panel). Above each violin are the counts for images above the probability threshold, formatted as $N_{\text{M1}}$-$N_{\text{M2}}$-$N_{\text{max(M1,M2)}}$. Below the $x$-axis labels of the right panel are the recovery rates for each grade given the combination of the two models.}
    \label{fig:violins}
\end{figure}

The performance of the individual classifiers M1 and M2 appear very similar in both training and deployment, as shown in Figure~\ref{fig:violins} (see also Figure~\ref{fig:auc-loss}). Performance of the models in ``training" specifically refers to model performance on the validation set, which is, as a reminder, a set of images not used to update model weights. Figure~\ref{fig:violins} shows violin plots for the distribution of probabilities from each model (M1 and M2) for objects, grouped by grade, in deployment (left panel) and training (right panel). Each violin shows the distribution of predictions from each model (M1: left, M2: right), above the respective 99.9th percentile probability determined in deployment. The black outline shows the distribution of the joint maximum probabilities following application of the joint probability threshold cut. In the right panel we apply the same deployment probability cuts to the training sample. In order to produce this plot we assign grades to images of strong lenses in the training sample in the same way that we assign grades in deployment (see $\S$~\ref{sec:inspection}). The horizontal black lines in each violin indicate the maximum, median, and minimum probabilities. 
We see a clear correlation between the model predicted probability and grade assigned via visual inspection for both the deployment and training samples. For each Grade A, B, C, and below (non-lens), the median probability monotonically decreases. To our knowledge, this is the first time that this trend has been reported in the literature.
The correlation has not been reported in the past, likely due to the fact that the necessity to inspect and grade objects in the training sample is only relevant when using real images of known strong lenses to train. It is important to note that no grade or quality indicators were factored into the training process. The ResNet has no insight into the quality of the system and thus this trend arises completely naturally, indicating that the ResNet assigns probability on a similar basis to that of human inspectors. Understanding this is critically important when considering future deployment and inspection strategies to recover the highest-quality strong lens candidates efficiently. 

Above each violin plot in Figure~\ref{fig:violins} are the counts for each grade from the individual models and the combination of models (M1-M2-combination). 
The combination of the models consistently results in a higher completeness than the individual models, although purity does decrease slightly. In training, the combination of the models results in an increase of completeness by 8\% and a decrease of purity by 3\% over either individual model. Leveraging differences in an ensemble of model predictions is a common strategy in recent searches for strong lenses \citep[e.g.,][]{shu2022a}. Introducing such variance into model predictions can prove to be highly successful in improving purity and completeness in deployment.
This could be done easily when training with large simulated image samples. However, when working with limited, real data sets as in this work, it may be necessary to utilize methods such as k-fold cross-validation \citep[][]{stone1974,geisser1975} training in the future in order to improve search yield and efficiency. 

Further analyzing training results, 
we can attempt to predict the fraction of non-lenses misclassified as lenses above the probability threshold \citep{huang2021a}, 
given as $(N_\text{l}~\times~r~\times~\frac{(1-p)}{p})~/~N_\text{nl}$.
$N_\text{l}(=232)$ and $N_\text{nl}(=10748)$ are the number of lenses and non-lenses in the validation set, $r(=0.625)$ is the recall, and $p(=0.954)$ is the precision. This results in an estimate of the fraction of non-lenses misclassified as lenses of 6.5$\times10^{-4}$. With the expectation that there is one lens in $\mathcal{O}(10^4)$ galaxies (or non-lenses), we expect one lens for every 6.5 non-lenses. In deployment, we find a rate of one lens for every 6.7 non-lenses (Grade A+B+C:non-lens, see Figure~\ref{fig:violins}), very similar to our expectation. 
However, if the rate of lensing deviates from one lens in $\mathcal{O}(10^4)$ galaxies, our estimates would not be consistent. If the rate of lensing is higher, we would find that our models do not perform as well in deployment as expected, given the training results.

However, we recognize that many of these images are likely to be false positives from the ResNet.
We see examples of ``super-human" classification by ResNet in E.~Silver et al. (submitted) which identifies such lenses in Hubble Space Telescope (HST) imaging. Point (2) is related to (1) in the sense that limitations of data may prohibit the identification of lens candidates in both an absolute and relative sense. The inherent limitation on image quality and/or depth dictate the size and brightness of the strong lensing systems that are discoverable \citep[see][]{collett2015a}. Additionally, the differences in depth and image quality across the three surveys included in this search (WHIGS, CFIS, and PS) impose a particular selection function on discovery that is not well understood. Investigation of this effect is outside the scope of this work. The final point, (3), artificially limits the potential discovery sample. In particular, imposing a $z$-band magnitude cut of 20 imposes a harsh limitation on the selection of strong lenses from the underlying population. 

\begin{figure}[H]
    \centering
    \includegraphics[width=0.98\textwidth]{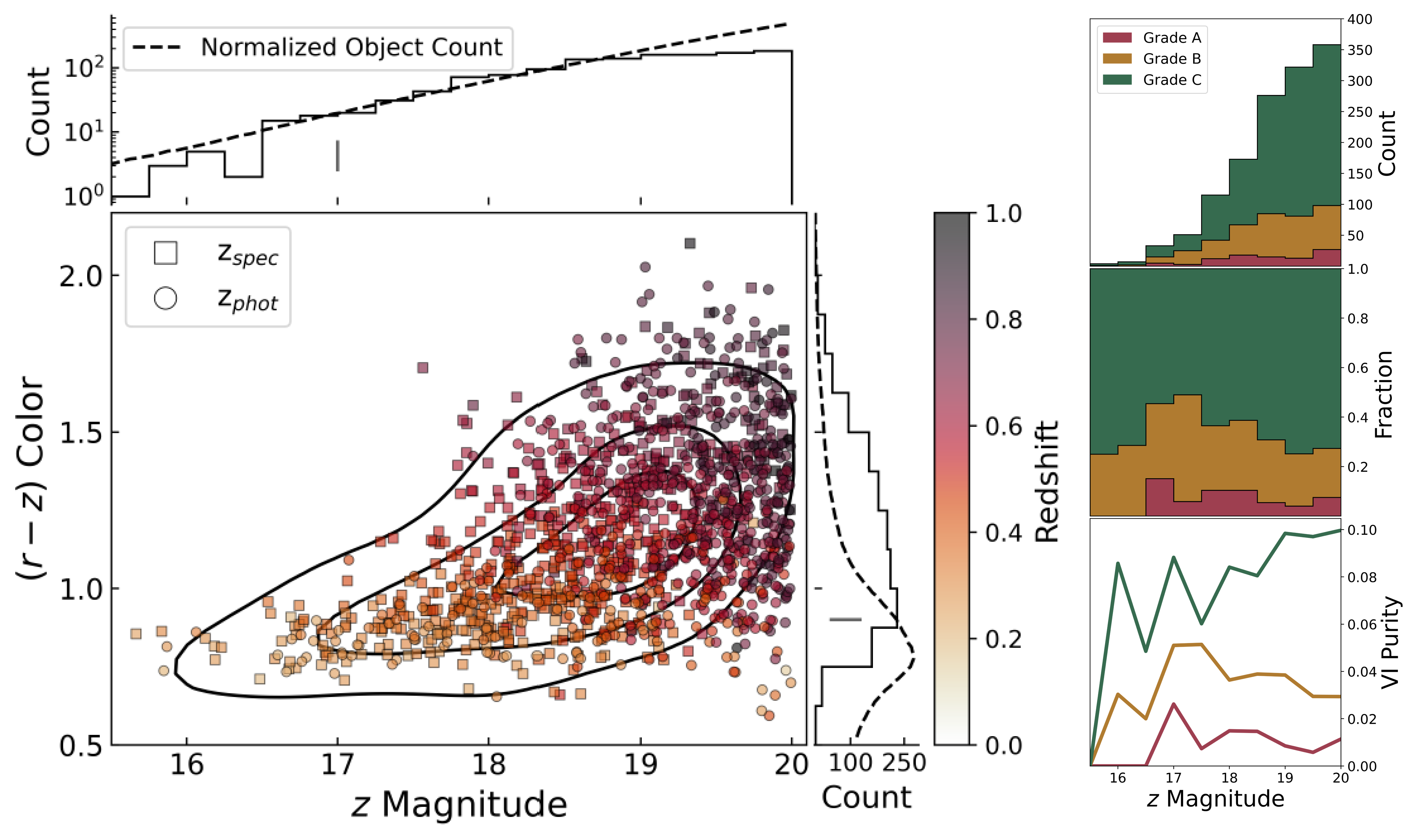}
    \caption{\footnotesize \textit{Left:} Scatter plot showing \textit{the Tractor} (\textit{r}-\textit{z}) color versus \textit{z} magnitude for lens galaxies of all \totalprelim candidates. The points are colored by redshift, where lenses with spectroscopic redshifts (SDSS or DESI) are indicated by squares and photometric redshifts as circles. The black contours show density of lenses in the training sample in the color-magnitude space generated with a kernel density estimator, showing contours which approximately correspond to the density percentiles of 68.3\%, 99.7\%, and 99.99994\%
    The top and right histograms in the margins correspond to the distribution of lens galaxy magnitude and color, respectively. Shown on both, as black dashed lines, are the distributions of all galaxies in \textit{the Tractor} object catalog (normalized at the location of the perpendicular black lines). \textit{Right:} Each panel shows information about candidates as a function of lens $z$-band magnitude, broken down by grade. The top panel shows the total counts, the middle panel shows the fractional representation of each grade, and the bottom panel shows the visual inspection purity.}  
    \label{fig:color-mag}
\end{figure}

Figure~\ref{fig:color-mag} further explores the broader implications of point (3) mentioned above, with respect to ResNet performance. In the left, primary panel of the figure, we show a scatter plot of lens ($r-z$) color versus $z$-band magnitude. The color of the markers indicate either the photometric or spectroscopic lens redshift (see $\S$~\ref{sec:spectro}). We also show the distribution of color and magnitudes in histograms in the margins and include the normalized object count from the deployment sample as the dashed lines. From the $z$-band magnitude histogram it is clear that the distribution is either peaking around 20th magnitude or has not yet peaked. This indicates that there are additional strong lens candidates that can be identified beyond this point. The black contours show the 1, 3, and 5$\sigma$ range for the training sample. The overall distribution of color and magnitude span a similar range as the new candidates reported in this work. This includes a very similar distribution of lens ($r-z$) color. However, the distribution of lens $z$ magnitudes is distinct in that there is a clear peak in the distribution, which occurs in the 18.75--19.25 magnitude bins. This indicates that the ResNet is capable of identifying lenses at fainter $z$ magnitudes, motivating a future search in fainter bins. 
\citet[][]{huang2021a} similarly shows that the distribution of reported lens candidates is considerably fainter than that of their training sample implying the ability to discover lenses in $z$ magnitude bins $>20$. 
We show additional motivation for searches in fainter $z$ magnitude bins through the comparison of $z$ magnitude distribution and the normalized object counts of the deployment sample. This can be used, not as an indicator of absolute completeness but of relative completeness across the magnitude range. Normalized at 17th magnitude, we observe exceptional relative completeness out to about 19th magnitude. Beyond that point we do not observe a dramatic decline in relative completeness, although it does worsen by a small amount, indicating minimal bias from the ResNet towards brighter lens galaxies. It is thus reasonable to expect that a ResNet trained on the images including the lenses reported in this work will be capable of discovering fainter lenses with a reasonably high completeness. 

The right-most panel of Figure~\ref{fig:color-mag} also shows evidence motivating future strong lens searches in UNIONS. Each of the three plots show characteristics of the sample of strong lens candidates reported in this work as a function of $z$ magnitude. The top plot shows the distribution of lens magnitude for each grade. There is a clear increase in the number of Grade C candidates and a more subtle increase in the Grade A and B candidates (though for the latter the Poisson noise is higher for each bin). While we expect the number of strong lensing systems in the underlying population to be increasing as a function of magnitude, many will become increasingly difficult to classify. Thus we expect the number of Grade C candidates to continue to increase at higher rates compared with Grade A and B candidates in even fainter $z$-band magnitude bins. The middle plot shows the fractional representation of each grade in each magnitude bin. Once again the decline in the fractional representation of Grade A and B candidates is not significantly diminishing in fainter magnitude bins. Note there is minimal representation of As and Bs in the $<17$th magnitude bins, likely indicating that these systems with brighter lens galaxies have already been discovered. In the bottom plot we show the visual inspection purity for each grade. While the purity of Grade C candidates continues to increase, the Grade A and B candidates, once again, do not diminish significantly in number. If the ResNet were to be retrained, including images of the candidates reported here, we expect the performance in the faintest magnitude bins to improve, also allowing for a relaxation of the magnitude cuts applied to the deployment sample.



\subsection{Brief Comparison with Similar Strong Lens Searches}
\label{sec:comparison}
\subsubsection{\citet{barroso2025a}}
\label{sec:AB25-comparison}
A recent UNIONS publication, \citet[][henceforth AB25]{barroso2025a}, reports the discovery of 132 new strong lens candidates, 82 of which are edge-on lens candidates, effectively doubling the number of such systems. They make use of the original version of CMU-Deeplens architecture \citep{lanusse2018a} implemented in \texttt{PyTorch}. As a reminder, we make use of a modified version of CMU-Deeplens implemented in \texttt{Tensorflow} in this work. In AB25, they make use of simulated $r$-band images of edge-on lens galaxy, strong lenses for training, and deploy on the CFIS $r$-band portion of UNIONS. They deploy on approximately 7 million extended sources ($17.0< r < 20.5$). Their search was conducted completely independently and in parallel with this work. 

With the specific focus on finding edge-on lenses, the overlap of the candidates in AB25 and those reported here are minimal, with only seven common candidates. These are shown in Figure~\ref{fig:barraso} and Table~\ref{tab:barraso}. Of these, only one, J1042+4320 (GLUE-160.7169+43.3421), is reported as an edge-on lens candidate. We additionally find that only three candidates reported in AB25 were included in our set of around 10000 ResNet recommendations, all of which received below Grade C quality. 
The remaining 122 new candidates reported in AB25 were either not in our deployment object catalog, which includes objects not in our search region (i.e., in areas without full \textit{gri} coverage or not in the NGC), or most commonly, they received a probability below our threshold for ResNet recommendations. As a reminder, we set a relatively high probability threshold in deployment, corresponding to the top 99.9\% of images. Coupled with the very specific selection function in AB25, with a ResNet trained on simulated edge-on lenses, we are not surprised by the low recovery rate of candidates. A total of 71 candidates are reported in AB25 (50 edge-on, 21 non edge-on) that fall under the probability threshold for our search. Of these there are 7 Grade A, 17 Grade B, and 47 Grade C candidates as reported in AB25. This corresponds to 50\%, 89\%, and 98\% of the totals for each graded candidate that overlaps with our deployment sample. 
The decreasing rate of recovery as a function of grade at the very least shows that while our absolute completeness may be relatively low, our ability to recover the highest grade candidates is respectable.
Despite the apparent success, however, this comparison highlights room for potential improvement in our ResNet performance and also in our deployment strategy.
Beyond this, the differences between our search and that of AB25 are large enough that, despite using the same ResNet architecture and deploying on the same data set, a more detailed comparison may not yield any additional insights. 
It does, however, underscore the importance of multiple, independent strong lens searches to drive up total completeness and maximize strong lens discoveries and the diversity in the known strong lensing population.

\vspace{0.5cm}
\begin{minipage}{\textwidth}
\makebox[\textwidth][c]{
  \includegraphics[keepaspectratio=true,width=0.98\linewidth]{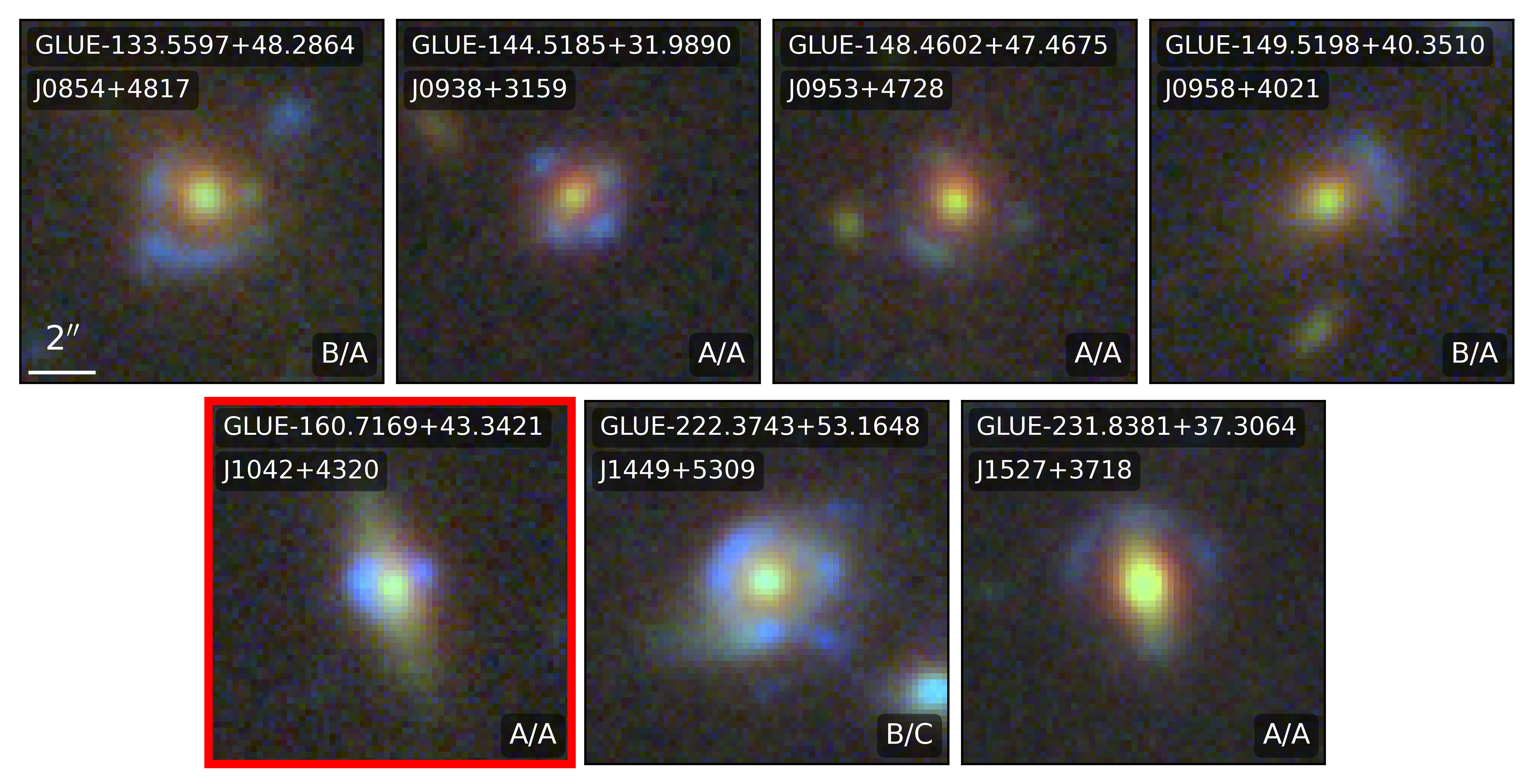}}
\captionof{figure}{\footnotesize Overlapping candidates reported in AB25 and this work. Each image panel is N up, and E to the left. System names from this work and AB25 are displayed in the upper left corner of each image panel along with system grades in the bottom right (this work/AB25).  The image outlined in red indicates the one candidate that is reported as an edge-on lens in AB25.}
\label{fig:barraso}
\end{minipage}
\vspace{0.25cm}

\begin{table}[h]
 \hspace{-1.cm}
 \centering
    \begin{tabular}{lcccc}
        \toprule
        \multicolumn{2}{c}{\textbf{AB25}} & \multicolumn{3}{c}{\textbf{This Work}} \\
        \midrule
        {Name} & {Grade} & {Name} & {Grade} & {Score} \\
        \midrule
        J0854+4817 & A & GLUE-133.5597+48.2864 & B & 3.5 \\
        J0953+4728 & A & GLUE-148.4602+47.4675 & A & 4 \\
        J0938+3159 & A & GLUE-144.5185+31.9890 & A & 4 \\
        J0958+4021 & A & GLUE-149.5198+40.3510 & B & 4 \\
        J1042+4320$^*$ & A & GLUE-160.7169+43.3421 & A & 3 \\
        J1449+5309 & C & GLUE-222.3743+53.1648 & B & 3 \\
        J1527+3718 & A & GLUE-231.8381+37.3064 & A & 3 \\
        \bottomrule
    \end{tabular}
    \caption{\footnotesize Overlapping lens candidates between AB25 and this work. The asterisk indicates the one overlapping system that is reported as an edge-on lens in AB25.}
    \label{tab:barraso}
    \vspace{-0.4cm}
\end{table}

\newpage

\subsubsection{\citet{storfer2024a}}
\label{sec:S24-comparison}
\citet[][S24]{storfer2024a} reports the largest catalog of new strong lens candidates from a single search in an imaging survey, at 1512. The search is conducted in the DESI LS DR9 using the exact same ResNet architecture as this work. The object catalog here is a subset of that used in S24, specifically, objects with $\delta>30^{\circ}$ (predominantly from the Beijing Arizona Sky Survey/Mayall \textit{z}-band Legacy Survey; BASS/MzLS). A number of candidates from the  BASS/MzLS region in S24 were used in the training sample in this work and the remainder were removed from the deployment sample (included in the removal of known lens candidates). For this reason we cannot make a direct comparison, including investigation of overlapping candidates, as was done for AB25. However, we can make a more general comparison including investigation of discovery rates and efficiency as well as the distribution of lens redshifts and magnitudes. 

The ResNet training in S24 was not optimized for the BASS/MzlS region. Rather, they optimized ResNet performance for the southern region of the DESI LS covered by Dark Energy Camera (DECam). Nonetheless, S24 reports the discovery of 162 strong lens candidates in 5000~deg$^2$ of survey area. In this work, we report \totalfinal new candidates in $4100$~deg$^2$. This translates to a discovery rate of one lens per 30 deg$^2$ and 3 deg$^2$ for the searches, respectively. S24 finds these candidates from inspection of 6500 ResNet recommendations, resulting in a reported efficiency of 1 lens for every 40 recommendations. In this work, from the $\sim10400$ recommendations, we find one lens for every 7.7 recommendations. Thus we are more efficient in area by a factor of about 10 but only more efficient in recommendations by a factor of 5. We believe that this, in addition to analysis of training results (see $\S$~\ref{sec:training_res}), indicates that our ResNet can be trained more effectively. In terms of success rate, S24 and this work find lens candidates at a rate of one lens per 50000 objects and 6000 objects, respectively.

S24 also reports the magnitudes and redshifts of their candidates. We find that the median $z$-band magnitude of lens candidates in S24 and this work are 18.91 and 19.01, respectively. This goes hand-in-hand with the lens galaxy redshifts, which are 0.54 and 0.58 respectively. The differences of 0.1 magnitudes and 0.04 in redshift are small, likely due to the use of the same deployment sample. Considering the improvement on depth and image quality from BASS/MzLS to UNIONS, it may be that in this search, we are capable of detecting strong lenses with smaller angular sizes as well as fainter lensed sources. Both types of new system are identified from the same parent population of galaxies and thus we see a similar distribution of lens galaxy characteristics compared with S24. Furthermore, our ability to probe systems with fainter $z$-band magnitudes at higher redshifts is inherently limited by our artificial magnitude cut.
Without spectroscopic and/or high-resolution imaging confirmation of a large sample of sources, it is difficult to form a conclusion about the difference in selection function for the respective surveys (BASS/MzLS and UNIONS) and the trained ResNets in from S24 and this work. However, it is clear that given the image quality, UNIONS allows for the discovery of systems with smaller Einstein radii. The depth of UNIONS also allows for the discovery of fainter lensed sources, specifically around the exact same population galaxies. Lastly, the total survey area of UNIONS allows for detection of an incredibly high volume of new strong lens candidates. Detection of smaller systems and fainter sources distinguishes the results presented in this work from those of S24. 
\\
\\
\subsubsection{Euclid: The Strong Lensing Discovery Engine}
\label{sec:SLDE-comparison}

Very recently, a suite of publications were released presenting and making use of the first public Quick Data Release (Q1) of Euclid observations \citep[][]{aussel2025a}. Among these publications, five involve investigations relating to the search for strong lenses by the Strong Lensing Discovery Engine \citep[SLDE;][]{SLDE-A2025a,SLDE-B2025a,SLDE-C2025a,SLDE-D2025a,SLDE-E2025a}, from which they report 497 strong lens candidates that they consider to be Grade A or B quality. They additionally include 1918 systems that receive a grade of C. The definition of Grade C in \citet[][]{SLDE-A2025a} is ``possible lens: shows lensing features but they can be explained without resorting to gravitational lensing”; they do not report these systems as strong lens ``candidates". For the purposes of the following comparison, we include systems reported with grades A, B, and C in SLDE. Note that despite using the same lettering scheme, the definition for grades and the corresponding expected quality for each is different between SLDE and this work. We consider our Grade Cs to be of high quality.
In SLDE, they conduct their searches in the Euclid Deep Field North (EDFN; North Ecliptic Pole), Euclid Deep Field South (EDFS; South Ecliptic Pole), and Euclid Deep Field Fornax (EDFF) totaling 63 deg$^2$. The EDFN region is completely contained within UNIONS and totals 22.9 deg$^2$, within which 183 strong lens candidates were reported in SLDE. In total, eight candidates (Grade A and B) as well as two systems with a grade of C reported in SLDE overlap with candidates reported in this work. These are shown in Figure~\ref{fig:SLDE} and Table~\ref{tab:SLDE}. 


While a comparison of selection function between these searches is important, it is outside the scope of this work. In deployment, SLDE achieves a much higher completeness than what is expected for this search due to the difference in image quality between Euclid and UNIONS. Therefore it is not unexpected that SLDE would successfully recover all candidates reported in this work that lie within the EDFN region. Of the 10 overlapping candidates between the two searches, six were assigned Grade C in this work. Of these, in SLDE, three were assigned Grade A, one Grade B, and two Grade C. This further emphasizes the quality of the Grade C candidates reported in this work and their potential to be confirmed via Euclid imaging. Assuming the EDFN region in UNIONS is a representative sample, we can make projections for all candidates assigned Grade C in this work. Four out of six of these Grade Cs were given grades of A or B in SLDE, thus we expect 66\%$\pm19\%$ or nearly 700$\pm$200 Grade C candidates could be confirmed (considered Grade A or B quality in SLDE) via Euclid. Continuing the assumption that the EDFN region of UNIONS is representative, we should expect about 58 ResNet recommendations. Indeed, we find 55 recommendations in the EDFN region, of which 45 were assigned as non-lenses in visual inspection in this work. In total, three systems which we consider to be non-lenses are reported in SLDE, two of Grade A or B quality and one Grade C quality (in SLDE; 3/45 nonlenses).
It is thus possible to expect that $6.7\%\pm3.7\%$ or $600\pm330$ systems classified as non-lenses (false negatives) in this search may be confirmed (considered grade A or B quality in SLDE) given Euclid imaging. These three non-lenses also all received probabilities $>0.99$ by one of our two ResNet models. This is in-line with expectations mentioned in $\S$~\ref{sec:deploy_perf} that some non-lenses may be be false negatives following human inspection. We acknowledge that these projections are made with a statistically small sample and are thus subject to large Poisson uncertainties.

\begin{minipage}{\textwidth}
  \vspace{0.3cm}
\makebox[\textwidth][c]{
  \includegraphics[keepaspectratio=true,width=0.8\linewidth]{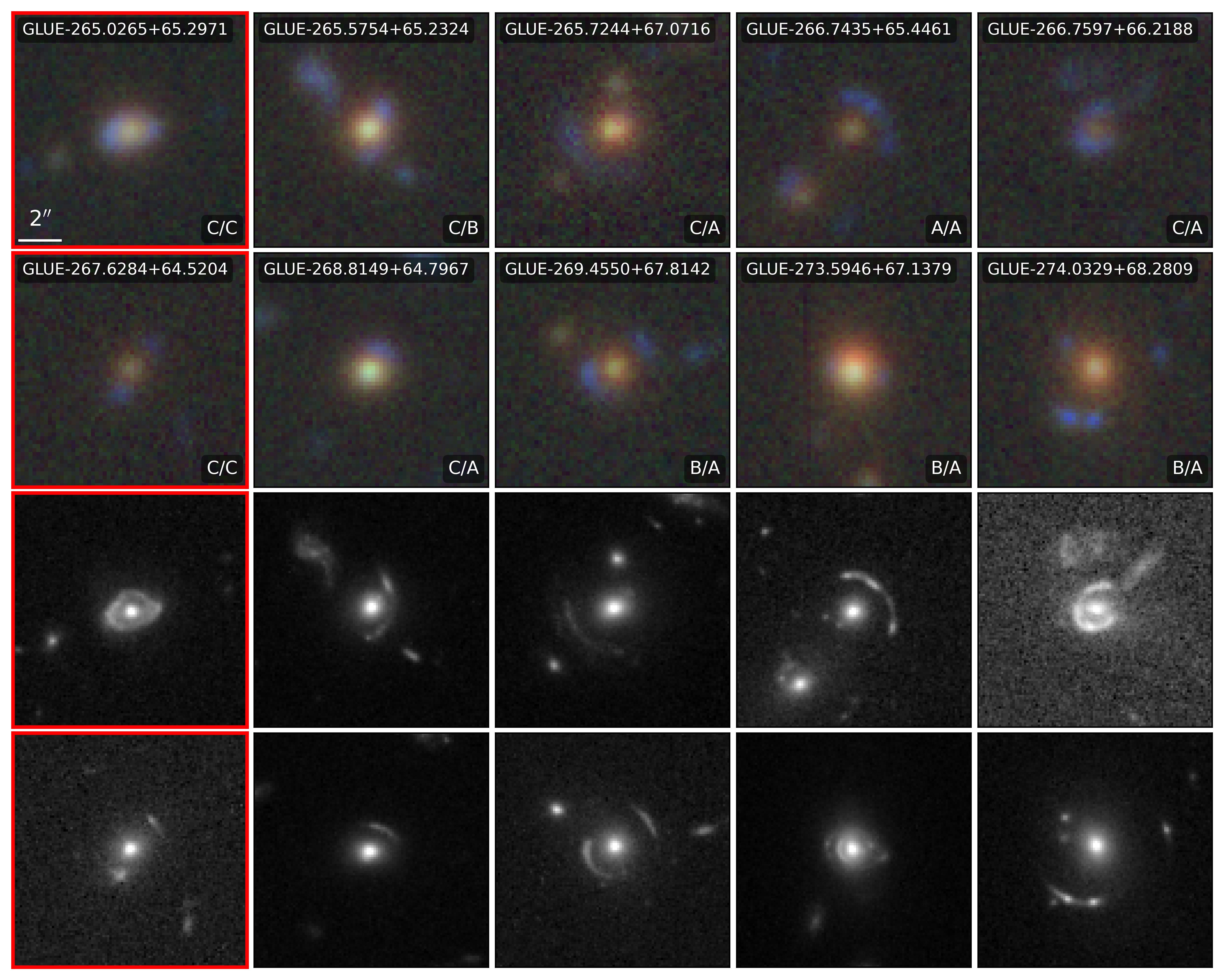}}
\captionof{figure}{\footnotesize Overlapping candidates reported in SLDE and this work. Each image panel has N up and E to the left. The top two rows display the UNIONS $gri$ color images of the candidates sorted by right ascension. System names from this work are displayed in the upper left corner of each UNIONS image panel along with system grades in the bottom right (this work/SLDE). The two bottom rows show the corresponding Euclid VIS images for the candidates. The panels outlined in red indicate candidates that received a grade of C from SLDE.}
\label{fig:SLDE}
\end{minipage}

\begin{table}[!ht]
    \centering
    \hspace{-1.8cm}
    \begin{tabular}{lccc}
    \hline
             & Grade & Grade & \\[-0.3em]
         Name& (SLDE)  & (This Work) & Score\\ \hline
        GLUE-265.7244+67.0716 & A & C & 2\\ 
        GLUE-266.7435+65.4461 & A & A & 2\\ 
        GLUE-266.7597+66.2188 & A & C & 3.5\\ 
        GLUE-268.8149+64.7967 & A & C & 2.5\\ 
        GLUE-269.4550+67.8142 & A & B & 2\\ 
        GLUE-273.5946+67.1379 & A & B & 3\\ 
        GLUE-274.0329+68.2809 & A & B & 3\\ 
        GLUE-265.5754+65.2324 & B & C & 3\\ \hline
        GLUE-265.0265+65.2971 & C & C & 2.5\\
        GLUE-267.6284+64.5204 & C & C & 2\\
        
    \hline
    \end{tabular}
    \caption{\footnotesize Overlapping lens candidates between the Euclid Strong Lensing Discovery Engine and this work. Note that in \citet{SLDE-A2025a} states that systems that received a grade of C are considered ``possible lens" and are separated by a horizontal line in the above table.}
    \label{tab:SLDE}
\vspace{-0.5cm}
\end{table}

\newpage

\subsection{Future Prospects with the Euclid Space Telescope}
\label{sec:euclid}
Of the 63 deg$^2$ of imaging included in Euclid Q1 in EDFN, EDFS and Fornax, projections from \citet[][]{collett2015a} show that the Euclid Q1 dataset may yield over 700 new strong lens candidates. The SLDE reports 497 candidates. With the Euclid coverage expected to grow to 2000 deg$^2$ following Data Release 1 (DR1), there will be thousands of new strong lenses discovered. UNIONS is and will continue to be highly complementary to the current and future Euclid data releases with respect to strong lens searches and beyond. Firstly, UNIONS covers a wide area that will not be observed by Euclid for years to come. Through continued searches in UNIONS, strong lenses are known \textit{a priori}, enabling preemptive spectroscopic follow-up, preliminary modeling, and potential monitoring for strongly lensed SNe. Secondly, with a single wide bandpass optical channel (VIS), Euclid will lack optical color information. Euclid does have near-infrared (NIR) $Y$, $J$, and $H$ band imaging capabilities via the Near Infrared Spectrometer and Photometer (NISP) instrument.
NISP has spatial resolution more similar to ground-based imaging but provides color information that may be useful for identification of a subset of strong lensing systems. However, we believe that the optical color information, which UNIONS provides, is more important for the majority of strong lens identification. Euclid and UNIONS have potentially great synergy in strong lens searches by combining imaging data from the surveys to make use of optical color (UNIONS), NIR color (NISP), and high-resolution optical imaging (VIS). This idea was originally suggested in \citet[][]{metcalf2018a} in reference to the Rubin Observatory (LSST) and Euclid data. 

\section{Conclusion} \label{sec:conc}
In this work we report the findings from a search for strong gravitational lenses in the 4100~deg$^2$ of the $gri$ band data from the Ultraviolet Near-Infrared Optical Northern Survey (UNIONS). We use an existing deep residual convolutional neural network \citep[ResNet,][]{huang2021a,storfer2024a} trained on real images of known strong lenses in the UNIONS footprint, as well as non-lenses. We deploy the trained ResNet on UNIONS images of 8 million galaxies brighter than 20th magnitude in the $z$ band selected from a catalog produced for the Dark Energy Spectroscopic Instrument Legacy Surveys. 
In total we find \totalfinal new candidates, of which \gradeAfinal are Grade A, \gradeBfinal Grade B, and \gradeCfinal Grade C. A total of 15 candidates were co-discovered by others: \citet{barroso2025a} (7) and \citet{SLDE-A2025a} (8). 
While we report a large number of new strong lens candidates in this work, we have also established a pipeline for strong lens discovery in UNIONS that shows promise for future successful searches.

Similar to \citet{huang2020a}, \citet{huang2021a}, and \citet{storfer2024a}, we use real images of lenses and non-lenses to train our ResNet. We exceed the highest purity reported in those works, finding one lens for every 7.7 recommendations compared with one lens for every 40 recommendations in the BASS/MzLS portion of the DESI LS \citep{storfer2024a}. This is despite conducting our search in perhaps the most heterogeneous single data set to date. Again, similar to the entirety of the DESI LS, the data from UNIONS used in this work is comprised of data from four telescopes at three observatories. However, unlike the DESI LS, the UNIONS dataset is not processed uniformly across filters presenting challenges in training which were addressed with image pre-processing. 

We find that the discoveries reported here are inherently limited by the $z$-band magnitude cut ($z < 20$). We show that not only is the ResNet capable of discovering fainter lenses than the training sample, but that it performs consistently well in most $z$-band magnitude bins, with only minimal declines in the faintest bins. Thus we are confident that future searches, including fainter lens galaxies, will reveal additional candidates at a respectable rate. These candidates will also likely be skewed to higher lens redshifts, increasing the sparsely populated space at $z_\text{d}\gtrsim0.8$

Of the candidates reported in this work, 580 have existing spectroscopic lens redshifts from SDSS DR18 (283) and DESI DR1 (297). All remaining candidates have photometric lens redshifts from \citet{zhou2020a} using DESI LS photometry. Despite not searching for explicitly high-redshift lenses, these systems have a median redshift of 0.58, 230 of which have $z_\text{d}>0.75$. An additional 15 new systems have been fully confirmed via SDSS (3) and DESI (12) through identification of a superposition of spectra at different redshifts in a single spectroscopic fiber. These were found through inspection of systems with large differences in purported lens photometric and spectroscopic redshifts where the photometric fit corresponds to the lens and spectroscopic fit to the source redshift. This methodology shows promise as a new and independent method for discovery of strong lenses. 

Moving forward, UNIONS will maintain its unique role in strong lens finding. Its combination of depth, breadth, and image quality make it one of the best data sets in the world to find strong lenses. 
While the Euclid Space Telescope will discover lenses at a higher rate than what is possible in UNIONS, the optical color information provided by UNIONS will continue to contribute to strong lens finding even after the completion of the 14,000 deg$^2$ Euclid survey. Upon completion, Euclid itself is projected to discover $\mathcal{O}(10^5)$ strong lenses, exceeding all discoveries from ground-based surveys combined. Even prior that point, identifying lenses in UNIONS holds tremendous value in informing spectroscopic follow-up, preliminary lens modeling, and even monitoring for strongly lensed SNe. Once observed by Euclid, these strong lenses can be confirmed via high resolution imaging and modeled immediately. Modern lens modeling pipelines which utilize graphics processing units (GPUs) such as GIGA-Lens \citep{gu2022a} are capable of (and necessary for) modeling the extraordinary number of strong lenses that will be discovered in the coming years. This includes discoveries from current and future imaging surveys such as UNIONS, the Vera C. Rubin Observatory Legacy Survey of Space and Time, Euclid, and the Nancy Grace Roman Space Telescope, as well as spectroscopic surveys including DESI and the 4-metre Multi-Object Spectroscopic Telescope \citep[4MOST;][]{dejong2019a}. These surveys will enable a breadth of strong lensing science from cosmology to galaxy evolution which will be made more robust from a large, representative population of strong lensing systems.
\section*{Acknowledgments}
We are honored and grateful for the opportunity of observing the Universe from Maunakea and Haleakala, which both have cultural, historical and natural significance in Hawaii. 

This work is based on data obtained as part of the Canada-France Imaging Survey, a CFHT large program of the National Research Council of Canada and the French Centre National de la Recherche Scientifique. Based on observations obtained with MegaPrime/MegaCam, a joint project of CFHT and CEA Saclay, at the Canada-France-Hawaii Telescope (CFHT), which is operated by the National Research Council (NRC) of Canada, the Institut National des Science de l’Univers (INSU) of the Centre National de la Recherche Scientifique (CNRS) of France, and the University of Hawai'i. 

The Pan-STARRS1 Survey (PS1) and the PS1 public science archive have been made possible through contributions by the Institute for Astronomy, the University of Hawai'i, the Pan-STARRS Project Office, the Max Planck Society and its participating institutes, the Max Planck Institute for Astronomy, Heidelberg, and the Max Planck Institute for Extraterrestrial Physics, Garching, The Johns Hopkins University, Durham University, the University of Edinburgh, the Queen’s University Belfast, the Harvard-Smithsonian Center for Astrophysics, the Las Cumbres Observatory Global Telescope Network Incorporated, the National Central University of Taiwan, the Space Telescope Science Institute, the National Aeronautics and Space Administration under grant No.~NNX08AR22G issued through the Planetary Science Division of the NASA Science Mission Directorate, the National Science Foundation grant No.~AST-1238877, the University of Maryland, Eotvos Lorand University (ELTE), the Los Alamos National Laboratory, and the Gordon and Betty Moore Foundation. Pan-STARRS is a project of the Institute for Astronomy of the University of Hawaii, and is supported by the NASA SSO Near Earth Observation Program under grants 80NSSC18K0971, NNX14AM74G, NNX12AR65G, NNX13AQ47G, NNX08AR22G, 80NSSC21K1572, and by the State of Hawaii.

Based on data collected at the Subaru Telescope and retrieved from the HSC data archive system, which is operated by Subaru Telescope and Astronomy Data Center at National Astronomical Observatory of Japan. The Hyper Suprime-Cam (HSC) collaboration includes the astronomical communities of Japan and Taiwan, and Princeton University. The HSC instrumentation and software were developed by the National Astronomical Observatory of Japan (NAOJ), the Kavli Institute for the Physics and Mathematics of the Universe (Kavli IPMU), the University of Tokyo, the High Energy Accelerator Research Organization (KEK), the Academia Sinica Institute for Astronomy and Astrophysics in Taiwan (ASIAA), and Princeton University. Funding was contributed by the FIRST program from Japanese Cabinet Office, the Ministry of Education, Culture, Sports, Science and Technology (MEXT), the Japan Society for the Promotion of Science (JSPS), Japan Science and Technology Agency (JST), the Toray Science Foundation, NAOJ, Kavli IPMU, KEK, ASIAA, and Princeton University. 

This research used the facilities of the Canadian Astronomy Data Centre operated by the National Research Council of Canada with the support of the Canadian Space Agency.

This research used the Canadian Advanced Network For Astronomy Research (CANFAR) operated in partnership by the Canadian Astronomy Data Centre and The Digital Research Alliance of Canada with support from the National Research Council of Canada, the Canadian Space Agency, CANARIE and the Canadian Foundation for Innovation.

The technical support and advanced computing resources from University of Hawaii Information Technology Services – Research Cyberinfrastructure, funded in part by the National Science Foundation CC* awards No.~2201428 and No.~2232862 are gratefully acknowledged.

C.J. Storfer is supported by Science Graduate Student Research (SCGSR) fellowship.
This material is based upon work supported by the U.S. Department of Energy, Office of Science, Office of Workforce Development for Teachers and Scientists, Office of Science Graduate Student Research (SCGSR) program. The SCGSR program is administered by the Oak Ridge Institute for Science and Education (ORISE) for the DOE. ORISE is managed by ORAU under contract number DE- SC0014664. All opinions expressed in this paper are the author’s and do not necessarily reflect the policies and views of DOE, ORAU, or ORISE.

X.H. acknowledges the Faculty Development Fund from the University of San Francisco.
\clearpage
\bibliography{dustarchive}



\end{document}